\documentclass[12pt]{iopart} 

\usepackage{amsfonts,amssymb,amsthm,amscd}
\usepackage{acronym,array,bm,ccaption,color,dcolumn}
\usepackage{epsfig,graphicx,nomencl,revsymb4-1,upgreek,url}
\usepackage{hyperref}
\hypersetup{colorlinks=true, linkcolor=blue, pdfauthor=Mohan Sarovar, pdftitle=}
\newcommand{\bra}[1]{\ensuremath{\left\langle{#1}\right\vert}}
\newcommand{\ket}[1]{\ensuremath{\left\vert{#1}\right\rangle}}

\renewcommand{\Re}{\mathbb{Re}}

\newcommand{\erf}[1]{Eq.~(\ref{#1})}

\newcommand{\beq}{\begin{equation}}
\newcommand{\eeq}{\end{equation}}
\newcommand{\bqa}{\begin{eqnarray}}
\newcommand{\eqa}{\end{eqnarray}}
\newcommand{\nn}{\nonumber}
\newcommand{\haqc}{\ensuremath{H_{\rm \tiny AQC}(t)}}

\newcommand{\haqcb}{\bar H_{\rm {AQC}}(t)}
\newcommand{\hc}{\ensuremath{H_{\rm C}}}
\newcommand{\hb}{\ensuremath{H_{\rm B}}}
\newcommand{\uc}{\ensuremath{\mathcal{U}_{\rm C}}}
\newcommand{\ub}{\ensuremath{\mathcal{U}_{\rm B}}}

\newcommand{\bnu}{{\boldsymbol{\nu}}}
\newcommand{\bmu}{{\boldsymbol{\mu}}}

\newcommand{\ie}{\emph{i.e.,}~}
\newcommand{\eg}{\emph{e.g.,}~}

\newcommand{\text}[1]{\textrm{#1}}

\begin{document}

\title[Error suppression and error correction in AQC: non-equilibrium dynamics]{Error suppression and error correction in adiabatic quantum computation: non-equilibrium dynamics}

\author{Mohan Sarovar}
\address{Scalable and Secure Systems Research, Sandia National Laboratories, MS 9158, 7011 East Avenue, Livermore, California 94550 USA}
\ead{mnsarov@sandia.gov}
\author{Kevin C.~Young}
\address{Scalable and Secure Systems Research, Sandia National Laboratories, MS 9158, 7011 East Avenue, Livermore, California 94550 USA}


\begin{abstract}
\noindent While adiabatic quantum computing (AQC) has some robustness to noise and decoherence it is widely believed that encoding, error suppression and error correction will be required to scale AQC to large problem sizes. Previous works have established at least two different techniques for error suppression in AQC. In this paper we derive a model for describing the dynamics of encoded AQC and show that previous constructions for error suppression can be unified with this dynamical model. In addition the model clarifies the mechanisms of error suppression and allow identification of its weaknesses. In the second half of the paper we utilize our description of non-equilibrium dynamics in encoded AQC to construct methods for error correction in AQC by cooling local degrees of freedom (qubits). While this is shown to be possible in principle, we also identify the key challenge to this approach: the requirement of high-weight Hamiltonians. Finally, we use our dynamical model to perform a simplified thermal stability analysis of concatenated-stabilizer-code encoded many-body systems for AQC or quantum memories.

This work is a companion paper to ``\textit{Error suppression and error correction in adiabatic quantum computation I: techniques and challenges}'' (Phys. Rev. X, 3, 041013 (2013)), which provides a quantum information perspective on the techniques and limitations of error suppression and correction in AQC. In this paper we couch the same results within a dynamical framework, which allows for detailed analysis of the non-equilibrium dynamics of error suppression and correction in encoded AQC. 
\end{abstract}

\maketitle

\tableofcontents

\section{Introduction}
\noindent Adiabatic quantum computing (AQC) is an alternative to the conventional circuit model for quantum computing that possess some distinct advantages. It maintains a many-body quantum system in its ground state while the Hamiltonian of the many-body system is morphed from a simple, typically non-interacting, form to a complex, connected form. The ground state of the final, complex Hamiltonian is designed to encode the solution to the problem being solved (\eg a satisfiability problem whose constraints are enforced by the Hamiltonian \cite{Farhi:2000tw,Farhi:2001ug}). During this evolution, energy relaxation and dephasing in the eigenbasis do not corrupt the computation and these are two reasons why AQC is believed to have some robustness to environmental fluctuations and noise. However, environmental fluctuations are typically local in space and for a general many-body system such local fluctuations may not result in only energy relaxation and depahsing in the eigenbasis. Thus the problem of evaluating the robustness of AQC is fundamentally linked to understanding the non-equilibrium dynamics of a many-body \textit{open} quantum system.

It was recognized in Refs.  \cite{Jordan:2006jb, Lid-2008} that one can gain some protection against external fluctuations (errors) by encoding the system state and AQC evolution in a quantum error correcting or detecting code. The properties of stabilizer codes \cite{Got-1997} allow one to suppress errors without effecting the adiabatic evolution by energetically penalizing them \cite{Jordan:2006jb} or inhibiting their action by dynamical decoupling \cite{Lid-2008, Quiroz:2012ea}. Both of these mechanisms are effective for error \textit{suppression} but perform no error correction. We refer to the suppression methods in Ref.  \cite{Jordan:2006jb} as energy gap protection (EGP) and the technique in Ref. \cite{Lid-2008} as dynamical decoupling (DD).

As in the conventional circuit-model of quantum computing error suppression techniques alone are insufficient for achieving fault-tolerant AQC. To have a fault-tolerant construction one requires a method for actively reducing the entropy of the encoded system such as error correction. However, it is a challenge to introduce error correction into AQC, which proceeds by Hamiltonian, continuous-time evolution. In contrast, circuit-model quantum computation has the advantage of being described by a sequence of discrete unitary gates within which error correction mechanisms (\eg syndrome measurement, correction rotations) can be embedded. 

In this work we view an encoded AQC evolution as a time-dependent many-body open quantum system, and the task of error correction as entropy reduction of this system. We ask the question: if the entropy reduction is performed by cooling local degrees of freedom (the notion of locality will be made more explicit below), can error correction be achieved in the AQC model of quantum computation? We find that error correction by local cooling can be achieved with a slightly modified AQC model, but at a heavy price. 

In the companion paper to this work  \cite{Young:2013} we present general constructions and arguments for the unification of EGP and DD, for the limitations of error suppression in AQC, and a framework for error correction by local cooling in AQC. In this work focus on the derivation of generalized master equations for describing the evolution of an encoded AQC system in the presence of environmental fluctuations. These equations allow one to rigorously simulate encoded AQC evolution in the presence decoherence and cooling, and obtain bounds on performance. In addition, the conditions required for error correction by local cooling in AQC are naturally revealed in the process of deriving the generalized master equations below. 

We emphasize that the error suppression and correction we are considering in this work are relevant for errors due to coupling to uncontrolled degrees of freedom. We do not consider errors that result from diabatic transitions resulting from evolution that is not slow enough (see Ref. \cite{Young:2013} for a complete discussion of the various failure modes in AQC). Such diabatic errors can be prevented by choosing conservative adiabatic speeds, by engineering the final Hamiltonian to increase energy gaps \cite{Somma:2013uc}, or by using prior knowledge of energy gaps to adapt the interpolation speed \cite{Roland2002,Nehrkorn:2011vo,Brif:2013vp}, although it is not known how to systematically do the latter in general large-scale problems.

Finally, we remark that in order to derive the dynamical equations used in this work we heavily exploit the structure of error correction codes. This is an illustration of a more general point: while understanding the dynamics of many-body \textit{open} quantum systems is extremely difficult in general, error correction code structure (or in other words, symmetries relevant to the system-environment coupling) can be exploited to make this problem more tractable. This highlights the utility of using quantum coding concepts to describe many-body quantum systems.

An outline of the remainder of the paper is: section \ref{sec:intro} presents the framework of stabilizer code encoded AQC and sets notation. Section \ref{sec:master} presents the derivation of a master equation capable of capturing the error suppression and correction dynamics of encoded AQC. Section \ref{sec:error_supp} manipulates this master equation to formalize the effects of error suppression by EGP and DD and quantify the performance of both. In particular, this section shows how both methods for error suppression are unified under the same dynamical picture. Section \ref{sec:error_corr} begins with a discussion of approaches to error correction in AQC and points out some challenges. Then we set out the Hamiltonian structure required for error correction by local cooling and follow this with a derivation of non-equilibrium dynamics of encoded, cooled, AQC. Section \ref{sec:ft} explores the properties required for long-term operation of an error corrected AQC system, thermal stability, and posits a possible definition of fault-tolerance in AQC. Finally, in section \ref{sec:disc} we conclude the paper with a summary of the results and commentary on the prospects for error correction and fault tolerance in AQC. 

\section{Encoded AQC} 
\label{sec:intro}
A general description of a quantum many-body system undergoing adiabatic evolution while coupled to an environment is given by the Hamiltonian:
$H(t) =  \haqc{} + \sum_{j=1}^{n_e} E_j\otimes B_j  + \hb.$
Here $H_{\rm AQC}$ acts on the many-body system and its time dependence implements the adiabatic evolution. In the following we will specialize to the many-body system most relevant to adiabatic quantum computing, a collection (of arbitrarily coupled) $n$ qubits. $\hb$ is the bath Hamiltonian, $B_j$ is a bath operator, and $E_j$ is a single qubit Pauli error operator ($j$ is not an index for the qubits but the set of errors; one qubit can have multiple error operators).

The system-bath interaction terms cause transitions from the adiabatically evolving ground state, and can result in a failed computation. The first step in protecting against these transitions is to encode the system in an error correcting or detecting stabilizer code \cite{Got-1997}. The code chosen must be one for which each $E_j$ in the system-bath interaction anti-commutes with at least one of the stabilizer generators.  This encoding will enlarge the system's Hilbert space by a factor of $2^{N_g}$, where $N_g$ is the number of stabilizer generators of the code.  The physical operators, $\sigma_x, \sigma_y, \sigma_z$ in $H_{\rm AQC}$ are replaced by the code's logical operators, $\bar X, \bar Y, \bar Z$, and the encoded Hamiltonian then becomes:
\begin{equation}
\label{eq:aqchambar}
	\bar H(t) =  \haqcb + \sum_{j=1}^{N_e} E_j\otimes B_j + \hb + \hc(t) 
\end{equation}
We have assumed that the system-bath interaction remains qualitatively the same after the encoding, but is extended to $N_e > n_e$ terms to correspond to the larger system size. We have also added a control Hamiltonian $\hc(t)$, which only acts on the system Hilbert space, and is required for error suppression and correction. The control Hamiltonian can take two forms. EGP \cite{Jordan:2006jb} chooses a static control Hamiltonian that is a sum of stabilizer generators $\hc^{\rm EGP}(t) = -\alpha \sum_{m=1}^{N_g} S_m$, with $\alpha >0$. States in the codespace are then eigenstates of $\hc$ with eigenvalue $-\alpha N_{g}$, but any state outside the codespace is subjected to an energy penalty. For dynamical decoupling based control, we sequentially apply the generators of the stabilizer group as unitary operators. In this case, the time-dependent control Hamiltonian is most easily written implicitly as the equivalent unitary that it generates:  
\begin{equation}
\uc^{\rm DD}(t) = \exp\left(-\frac{i}{\hbar}\int_0^t \hc^{\rm DD}(t') dt' \right) =  \prod_{j=0}^{K(t)} S_{\mathbf{n}_j}
\end{equation}
The stabilizers are applied in a particular order, given by the vector $\mathbf{n}$, at times given by $K(t)\in \mathbb{Z}$, so that at time $t$, the last operator applied to the system was $S_{\mathbf{n}_{K(t)}}$. These two control Hamiltonians can be viewed as two extremes of general control Hamiltonians where the stabilizer generators are applied time-dependently; see Ref. \cite{Young:2013} for a complete discussion of this. In the following we shall see that the DD control Hamiltonian is useful for error suppression, while the EGP control Hamiltonian is suitable for error suppression and error correction.

For the following discussion it will be useful to define several frames of reference. In order to do this, first define $\mathcal{U}_n(t_1,t_2) = {\rm exp}_+(-\frac{i}{\hbar}\int_{t_1}^{t_2} ds H_n(s))$, for $n \in \{{\rm B}, {\rm C}\}$ and $\mathcal{U}_{\rm AQC}(t_1,t_2) = {\rm exp}_+(-\frac{i}{\hbar}\int_{t_1}^{t_2} ds \bar{H}_{\rm AQC}(s))$  ($+$ denotes positive time ordering). For convenience, $\mathcal{U}_n(t) \equiv \mathcal{U}_n(t,0) ~\forall n$. Note that because of the commutation properties of the encoded and control Hamiltonians all of these unitaries commute with each other. The following notation is used for an operator $A$ in an interaction frame with respect to the control: $\tilde{A}(t) \equiv \mathcal{U}^\dagger_{\rm C}(t)\mathcal{U}^\dagger_{\rm B}(t) A~ \mathcal{U}_{\rm C}(t)\mathcal{U}_{\rm B}(t)$, which is typically called the \textit{toggling frame}. Evolution of states in this frame is generated by the toggling frame Hamiltonian: $\tilde{H}(t) \equiv\uc^\dagger(t)\ub^\dagger(t)\left(\bar H(t) -\hc -\hb \right)\ub(t)\uc(t)$.

It is shown in Ref. \cite{Young:2013} that the encoded AQC Hamiltonian in the toggling frame looks very similar for the two control scenarios, DD and EGP. Specifically, 
\begin{eqnarray}
	\tilde{H}_{\rm DD/EGP}
		&= \bar H_{\rm AQC}(t) + \sum_{j=1}^{N_e} \tilde E_j^{\rm DD/EGP}(t) \otimes \tilde B_j(t)
\end{eqnarray}
The form of the error operators in the toggling frame for the two control scenarios is \cite{Young:2013}:
\begin{eqnarray}
\label{eq:ejs}
	\tilde E_j^{\rm DD}(t) &=&(-1)^{p(t)} E_j \label{eq:ejdd}\\
	\tilde E_j^{\rm{EGP}}(t) &=&  e^{\left(-\frac{i}{\hbar} 2 \alpha t \sum_{\{S_m,E_j\}=0} S_m \right)} E_j, \label{eq:ejegp}
\end{eqnarray}
where $p(t) = 0$ if $[E_j,\uc^{\rm DD}(t)]=0$ and $p(t) = 1$ if $\{E_j,\uc^{\rm DD}(t)\}=0$. $p(t)$ encodes whether the last DD pulse applied commuted or anti-commuted with the error $E_j$ (note that since $\mathcal{U}_{\rm C}^{\rm DD}$ is always a member of the stabilizer group, $E_j$ must either commute or anti-commute with it at all times). An effective DD cycle is one that causes $p(t)$ to rapidly alternate between $+1$ and $-1$ in succession and thus the system-environment coupling is modulated by a rapidly oscillating function of $t$. Similarly, the sum in the exponential of \erf{eq:ejegp} is taken over all stabilizer generators $S_m$ that anti-commute with the error operator $E_j$. The modulation of the error operator in the EGP case is operator-valued, but the action on states in the code-space is very similar for both control scenarios, a fact we will exploit below.

\section{Dynamical master equation for encoded AQC} 
\label{sec:master}

In this section we will formalize the dynamics of error suppression and correction in encoded AQC by deriving a master equation describing effective encoded adiabatic evolution when the qubits are coupled to uncontrolled degrees of freedom. By employing fewer approximations than in the derivation of the conventional Lindblad master equation this reduced dynamics is able to capture the modification of system-environment coupling, and hence decoherence, by controls such as dynamical decoupling or energy penalty terms.

We begin with the full Hamiltonian in \erf{eq:aqchambar}, and assume that the system-environment coupling is weak compared to the other terms in this Hamiltonian, and that the encoding has been chosen such that each $E_j$ is a detectable error. Define an interaction frame with respect to $\bar{H}_{\rm AQC}(t)$, $H_{\rm C}(t)$ and $H_{\rm B}$ as: $\breve{A}(t) \equiv \mathcal{U}^\dagger(t,0) ~A~ \mathcal{U}(t,0)$, where
\begin{equation}
\mathcal{U}(t,0) = e_+^{-\frac{i}{\hbar} \int_0^t ds \bar{H}_{\rm AQC}(s) + H_C(s) + H_B}
\label{eq:int_u}
\end{equation}
A particularly important property of these Hamiltonians, which we will utilize later, is that they all commute. That is, $[\bar{H}_{\rm AQC}(s), H_{\rm C}(s')]=0 ~~~ \forall s,s'$, because $\bar{H}_{\rm AQC}$ only contains logical operators and $H_{\rm C}$ only contains stabilizer terms \cite{Got-1997}. And obviously $H_{\rm B}$ commutes with the other two terms. This property implies that this interaction frame transformation factors into: $\mathcal{U}(t,0) = \prod_{n={\rm AQC,C,B}} \mathcal{U}_n(t,0)$. 

Let $\varrho$ be the combined density matrix of system and environment, \ie a normalized trace-class operator in $\mathcal{H}_{\rm sys}\otimes \mathcal{H}_{\rm env}$. By substituting a formal solution to the von-Neumann equation we get the following dynamical equation for the combined density matrix in the interaction frame \cite{Bre.Pet-2002}:
\begin{equation}
\frac{{\rm d}\breve{\varrho}(t)}{{\rm d}t} = -\frac{i}{\hbar}[ \breve{H}_I(t), \breve{\varrho}(0)] - \frac{1}{\hbar^2}\int_0^t [\breve{H}_I(t), [\breve{H}_I(s), \breve{\varrho}(s)]] ds,
\end{equation}
where $H_I = \sum_j E_j \otimes B_j$ is the interaction Hamiltonian. We will assume that the weak system-environment coupling does not perturb the environment from its equilibrium state at timescales that we resolve, and hence $\breve{\varrho}(s) \approx \breve{\rho}(s) \otimes \sigma_{\rm eq}$, a tensor product of the system density matrix $\breve{\rho}(s) \equiv \tr_{\rm env} \{\breve{\varrho}(s)\}$, and the environmental equilibrium density matrix. This allows the derivation a time-convolution master equation for the system density matrix \cite{Bre.Pet-2002}:
\begin{eqnarray}
\frac{{\rm d}\breve{\rho}(t)}{{\rm d}t} &= \tr_{\rm env}\{\frac{{\rm d}\breve{\varrho}(t)}{{\rm d}t} \} \nonumber \\
	&= - \frac{1}{\hbar^2}\sum_{j,k} \int_0^t ds C_{kj}(t,s) \breve{E}_k(t)\breve{E}_j(s)\breve{\rho}(s) - C_{jk}(s,t) \breve{E}_k(t) \breve{\rho}(s) \breve{E}_j(s) \nonumber \\
	& ~~~~~~~~~~~~~ - C_{kj}(t,s) \breve{E}_j(s) \breve{\rho}(s) \breve{E}_k(t) + C_{jk}(s,t) \breve{\rho}(s) \breve{E}_j(s)\breve{E}_k(t)
\end{eqnarray}
with $C_{kj}(t,s) \equiv \tr_{\rm env}\{\breve{B}_k(t)\breve{B}_j(s)\sigma_{\rm eq}\}$ being the quantum correlation function of the environment. To obtain this expression we have assumed that $\tr_{\rm env}\{ \breve B_j(t) \sigma_{\rm eq}\} = 0 ~~\forall j$ -- \ie the average interaction force on the bath equilibrium state is zero. We further assume that the environment is stationary, implying that this correlation function is only dependent on the time difference $\tau = t-s$. This simplifies the master equation to:
\begin{eqnarray}
\frac{{\rm d}\breve{\rho}(t)}{{\rm d}t} &= - \frac{1}{\hbar^2} \sum_{j,k} \int_0^t d\tau C_{kj}(\tau) \breve{E}_k(t)\breve{E}_j(t-\tau)\breve{\rho}(t-\tau) \nonumber \\
	& ~~~~~~~~~~~~~ - C_{kj}^*(\tau) \breve{E}_k(t) \breve{\rho}(t-\tau) \breve{E}_j(t-\tau) \nonumber \\
	& ~~~~~~~~~~~~~ - C_{kj}(\tau) \breve{E}_j(t-\tau) \breve{\rho}(t-\tau) \breve{E}_k(t) \nonumber \\
	& ~~~~~~~~~~~~~ + C_{kj}^*(\tau) \breve{\rho}(t-\tau) \breve{E}_j(t-\tau)\breve{E}_k(t)
\label{eq:conv_me}
\end{eqnarray}
The final approximation we make is sometimes referred to as the \textit{first Markov approximation} \cite{Sch-2007} and replaces $\breve{\rho}(t-\tau)$ with $\breve{\rho}(t)$ in the integrals above. This amounts to assuming that the change in the system state (in the interaction frame, and therefore due to the weak system-environment coupling) is negligible on the timescale set by the decay of the environment correlation function. Therefore this formalism is valid for fast-relaxing or weakly-coupled environments. In the following we will restrict out analysis to uncorrelated environments for the system qubits, that is $C_{kj}(\tau) = \delta_{kj}C_j(\tau)$. The analysis that follows can be generalized to correlated environments but we will not do so here. 

We rewrite this resulting master equation in an interaction frame with respect to the control Hamiltonian only (the toggling frame): $\tilde{A}(t) = \mathcal{U}^\dagger_{\rm C}(t,0) ~A~\mathcal{U}_{\rm C}(t,0)$, for $A \in \mathcal{H}_{\rm sys}$. The transformation required to move into this frame is particularly easy in this case because as noted above the stabilizer properties result in a factoring of the full interaction frame transformation unitary (\erf{eq:int_u}). In the toggling frame:
\begin{eqnarray}
\frac{{\rm d}\tilde{\rho}(t)}{{\rm d}t} &= -\frac{i}{\hbar}[\bar{H}_{\rm AQC}(t), \tilde{\rho}(t)] \nonumber \\
	&- \frac{1}{\hbar^2} \sum_j \int_0^t d\tau C_j(\tau) \tilde{E}_j(t) \tilde{\Xi}_j(t,\tau)\tilde{\rho}(t)  - C_j^*(\tau) \tilde{E}_j(t) \tilde{\rho}(t) \tilde{\Xi}_j(t,\tau) \nonumber \\
	& ~~~~~~~~~~~~~ - C_j(\tau) \tilde{\Xi}_j(t,\tau) \tilde{\rho}(t) \tilde{E}_j(t) + C_j^*(\tau) \tilde{\rho}(t) \tilde{\Xi}_j(t,\tau)\breve{E}_j(t)
\label{eq:time-local_me}
\end{eqnarray}
where $\tilde{\Xi}_j(t,\tau) \equiv \mathcal{U}_{\rm AQC}(t,t-\tau)~\tilde{E}_j(t-\tau)~\mathcal{U}^\dagger_{\rm AQC}(t,t-\tau)$. 

We will mostly work with this time-local master equation with time-dependent dissipation kernels \cite{Bre.Pet-2002} in what follows. However, it is possible to also make the second Markov approximation \cite{Sch-2007} here and set the upper limit of the integrals above to $\infty$. This typically results in an equation of motion with no $t$-dependance on the incoherent transition rates. Physically, this approximation implies that the bath correlation functions decay so quickly that the time-dependence of the error operators (in the interaction frame) are not resolved by the integrals. Finally, the most drastic approximation replaces the correlation function with a delta function in time, which results in a temperature-independent master equation. We will not make this last approximation is this work since it results in evolution where the control Hamiltonian cannot influence the dissipation and decoherence operators directly, which is counter to any error suppression scheme. 

\section{Error suppression in AQC}
\label{sec:error_supp}
In order to understand the effects of error suppression on encoded AQC dynamics we begin by quantifying the population preserved in the codespace (the no-error subspace). We define $\mathbf{P}=\frac{1}{2^{N_g}} \prod_{m=1}^{N_g}(\mathbf{I} + S_m)$ as the projector onto the codespace, $P_0(t)$ as the codespace population at time $t$, and $\mathbf{Q} = \mathbf{I}-\mathbf{P}$. Then the change in the codespace population is $\frac{{\rm d}P_0}{{\rm d}t} = \tr\{\mathbf{P}\frac{{\rm d}\tilde{\rho}}{{\rm d}t}\mathbf{P}\}$. To evaluate this quantity, we will first insert identities in the form $\mathbf{P+Q}$ around $\tilde{\rho}(t)$, resulting in:

\begin{eqnarray}
\fl \frac{{\rm d}P_0(t)}{{\rm d}t} = - \tr \bigg\{ \sum_j \frac{1}{\hbar^2} \int_0^t d\tau 
		C_j(\tau) \bigg[ \mathbf{P}\tilde{E}_j(t) \tilde{\Xi}_j(t,\tau) \mathbf{P}\tilde{\rho}(t)\mathbf{P} + \mathbf{P}\tilde{E}_j(t) \tilde{\Xi}_j(t,\tau) \mathbf{Q}\tilde{\rho}(t)\mathbf{P} \bigg] \nonumber \\
	-C_j^*(\tau) \bigg[ \mathbf{P} \tilde{E}_j(t) \mathbf{Q} \tilde{\rho}(t)  \mathbf{Q} \tilde{\Xi}_j(t,\tau) \mathbf{P}\bigg] - C_j(\tau) \bigg[ \mathbf{P} \tilde{\Xi}_j(t,\tau) \mathbf{Q}  \tilde{\rho}(t) \mathbf{Q}  \tilde{E}_j(t) \mathbf{P} \bigg] \nonumber \\
	+C_j^*(\tau) \bigg[\mathbf{P} \tilde{\rho}(t) \mathbf{P} \tilde{\Xi}_j(t,\tau)\tilde{E}_j(t) \mathbf{P} + \mathbf{P} \tilde{\rho}(t) \mathbf{Q} \tilde{\Xi}_j(t,\tau)\tilde{E}_j(t) \mathbf{P} \bigg] \bigg\}
\end{eqnarray}
where we have used the identities: $\mathbf{P}\mathbf{Q} = 0$, $\mathbf{P} \tilde{E}_j \mathbf{P} = 0 ~~\forall j$, and $\mathbf{P} \tilde{\Xi}_j(t, \tau) \mathbf{P} = 0 ~~ \forall j$. The first of these is by definition and the others follow from the properties of the Hamiltonian and error operators -- \ie $\bar{H}_{\rm AQC}(s)$ and $H_{\rm C}(s)$ cannot move states between the subspaces projected onto by $\mathbf{P}$ and $\mathbf{Q}$, and $E_j$ applied to any state in $\mathbf{P}$ results in a state in $\mathbf{Q}$. The term $\tr\{ \mathbf{P}\tilde{E}_j(t) \tilde{\Xi}_j(t,\tau)\mathbf{Q}\tilde{\rho}(t)\mathbf{P}\}$ and its conjugate also evaluate to zero although it is slightly more involved to see why. The reason is that $\tilde{E}_j(t) \tilde{\Xi}_j(t,\tau) = \mathcal{U}^\dagger_{\rm C}(t,0)E_j\mathcal{U}_{\rm C}(t,0)\mathcal{U}_{\rm AQC}(t,t-\tau) \mathcal{U}^\dagger_{\rm C}(t-\tau,0)E_j\mathcal{U}_{\rm C}(t-\tau,0)~\mathcal{U}^\dagger_{\rm AQC}(t,t-\tau)$ contains two applications of $E_j$ interleaved with unitary evolution that does not connect different stabilizer syndrome subspaces and hence cannot connect $\mathbf{P}$ and $\mathbf{Q}$ subspaces. Hence, this master equation simplifies to:
\begin{eqnarray}
\fl \frac{{\rm d}P_0(t)}{{\rm d}t} = - \tr \bigg\{ \sum_j \frac{1}{\hbar^2} \int_0^t d\tau 
		C_j(\tau) \bigg[ \mathbf{P}\tilde{E}_j(t) \tilde{\Xi}_j(t,\tau) \mathbf{P}\tilde{\rho}(t)\mathbf{P} \bigg]  + C_j^*(\tau) \bigg[\mathbf{P} \tilde{\rho}(t) \mathbf{P} \tilde{\Xi}_j(t,\tau)\tilde{E}_j(t) \mathbf{P} \bigg] \nonumber \\
	-C_j^*(\tau) \bigg[ \mathbf{P} \tilde{E}_j(t) \mathbf{Q} \tilde{\rho}(t)  \mathbf{Q} \tilde{\Xi}_j(t,\tau) \mathbf{P}\bigg] - C_j(\tau) \bigg[ \mathbf{P} \tilde{\Xi}_j(t,\tau) \mathbf{Q}  \tilde{\rho}(t) \mathbf{Q}  \tilde{E}_j(t) \mathbf{P} \bigg] \bigg\} \nn
\end{eqnarray}

At this point we employ a critical property of the control Hamiltonian: that it modulates the system-environment interaction. Using the expressions for toggling frame error operators in \erf{eq:ejdd} and \erf{eq:ejegp} allows us to simplify the equation of motion for codespace population to:
\begin{eqnarray}
\label{eq:p0_noapprox}
\frac{{\rm d}P_0(t)}{{\rm d}t} =  \frac{2}{\hbar^2}\sum_{j=1}^{N_e} \int_0^t d\tau 
	&\Re \bigg\{ C_j(\tau) m_j(t,\tau) \tr \bigg[ \mathbf{P} \hat{\Xi}_j(t,\tau) \mathbf{Q}  \tilde{\rho}(t) \mathbf{Q} E_j \mathbf{P} \bigg] \bigg\} \nonumber \\ 
	&-\Re \bigg\{ C_j(\tau) m^*_j(t,\tau) \tr \bigg[ \mathbf{P}E_j \hat{\Xi}_j(t,\tau) \mathbf{P}\tilde{\rho}(t)\mathbf{P} \bigg]  \bigg\}
\end{eqnarray}
where $\hat{\Xi}(t,\tau) \equiv \mathcal{U}_{\rm AQC}(t,t-\tau)E_j~\mathcal{U}^\dagger_{\rm AQC}(t,t-\tau)$, , and $m_j(t,\tau)$ is a modulation function that results from the control. It captures the control influence on the dissipation and decoherence. For the two control scenarios of EGP and DD, the modulation functions take the form:
\begin{eqnarray}
m^{\rm EGP}_j(t,\tau) &= e^{\frac{i}{\hbar}2 \alpha \tau w_j} \label{eq:EGP_mod}\\
m^{\rm DD}_j(t,\tau) &= (-1)^{p(t)-p(t-\tau)}
\label{eq:DD_mod}
\end{eqnarray}
where $w_j$ is the number of stabilizer terms in the EGP penalty Hamiltonian that anti-commute with error $E_j$. $p(t)$ is the DD coefficient defined above. To write the modulation function for EGP we have exploited the property 
\begin{eqnarray}
e^{\left(\frac{i}{\hbar} 2 \alpha \tau \sum_{\{S_m,E_j\}=0} S_m \right)} \mathbf{P} = e^{\frac{i}{\hbar}2 \alpha \tau w_j } \mathbf{P}
\end{eqnarray}
which follows from the fact that all states in the codespace are eigenvalue $+1$ eigenstates of the stabilizers. These modulation functions are analogous to the \textit{filter functions} derived for describing dynamical decoupling for pure dephasing dynamics \cite{deSousa09}.

The modulation functions given in \erf{eq:EGP_mod}-(\ref{eq:DD_mod}) display some degree of asymmetry between the EGP and DD error suppression techniques because while $m^{\rm DD}$ depends on times $t$ and $\tau$, $m^{\rm EGP}$ only depends on time $\tau$. This is only because we have restricted ourselves to the case of constant, uniform energy penalty $\alpha$. As detailed in Ref. \cite{Young:2013}, a more general formulation of EGP would allow for $\alpha$ to be time dependent: $\hc^{{\rm EGP}}(t) = - \sum_{m=1}^{N_g} \alpha_m(t) S_m$. In this case,
\begin{equation}
m^{\rm EGP}_j(t,\tau) = (e^{\frac{i}{\hbar}2})^{\chi(t)-\chi(t-\tau)}
\end{equation} 
with $\chi(t) \equiv -\sum_{\{S_m,E_j\}=0} \int_0^{t} ds ~ \alpha_m(s)$. Comparing this with \erf{eq:DD_mod}, we see that in this more general formulation the similarity between DD and EGP is even more evident; they both modulate the dissipation kernels defining the leakage from the codespace (DD with a square pulse and EGP with a smooth oscillating function).

Since the correlation function, $C_j(\tau)$, decays with $\tau$ the value of the integrands in \erf{eq:p0_noapprox} at small values of $\tau$ are the most important. If we assume that the adiabatic interpolation, $H_{\rm AQC}$, varies slowly with respect to time, we can approximate
\begin{eqnarray}
\mathcal{U}_{\rm AQC}(t,t-\tau) = e_+^{-\frac{i}{\hbar}\int_{t-\tau}^t ds \bar H_{\rm AQC}(s)} \approx e^{-\frac{i}{\hbar} \tau \bar H_{\rm AQC}(t)}
\label{eq:aqc_approx} 
\end{eqnarray}
and hence approximate $\hat{\Xi}(t,\tau) \approx \Xi(t,\tau) \equiv e^{-\frac{i}{\hbar}\tau \bar H_{\rm AQC}(t)} E_j~e^{\frac{i}{\hbar}\tau \bar H_{\rm AQC}(t)}$ in \erf{eq:p0_noapprox}. Thus the final form of the population master equation is:

\begin{eqnarray}
\label{eq:pc_mastereq}
\frac{{\rm d}P_0(t)}{{\rm d}t} \approx  \frac{2}{\hbar^2} \sum_{j=1}^{N_e} \int_0^t d\tau 
	&\Re \bigg\{ C_j(\tau) m_j(t,\tau) \tr \bigg[ \mathbf{P} \Xi_j(t,\tau) \mathbf{Q}  \tilde{\rho}(t) \mathbf{Q} E_j \mathbf{P} \bigg] \bigg\} \nonumber \\ 
	&-\Re \bigg\{ C_j(\tau) m^*_j(t,\tau) \tr \bigg[ \mathbf{P}E_j \Xi_j(t,\tau) \mathbf{P}\tilde{\rho}(t)\mathbf{P} \bigg]  \bigg\}
\end{eqnarray}

\subsection{$H_{\rm AQC}=0$}
\label{sec:haqc_eq_0}
To see the effects of the control Hamiltonian even more clearly, we consider this dynamical equation in the absence of the adiabatic evolution (\ie when $H_{\rm AQC}=0$). Then the traces in this equation simplify further since $\tr\{E_j \Xi_j(t,\tau) \mathbf{P}\tilde{\rho}(t)\mathbf{P}\} \rightarrow \tr\{\mathbf{P}\tilde{\rho}(t)\mathbf{P}\}$, and $\tr\{E_j \mathbf{P} \Xi_j(t,\tau) \mathbf{Q}  \tilde{\rho}(t) \mathbf{Q}\} \rightarrow \tr\{\mathbf{Q}_1\tilde{\rho}(t)\mathbf{Q}_1\}$, where $\mathbf{Q}_1$ is a projector onto the subspace of $\mathbf{Q}$ that contains states one error away from the codespace. This simplification allows the derivation of a classical master/rate equation for the codespace population:
\begin{eqnarray}
\frac{{\rm d}P_0(t)}{{\rm d}t} \bigg|_{H_{\rm AQC}=0} &\approx  \sum_j r^+_j(t) P_{1}(t)  - \sum_j  r^-_j(t) P_0(t)
\label{eq:pop_eqn}
\end{eqnarray}
with 
\begin{eqnarray}
r^+_j(t) &\equiv \frac{2}{\hbar^2} \Re \big\{ \int_0^t d\tau C_j(\tau) m_j(t,\tau) \big\} \nonumber \\
r^-_j(t) &\equiv \frac{2}{\hbar^2} \Re \big\{ \int_0^t d\tau C_j(\tau) m^*_j(t,\tau) \big\},
\label{eq:r_rates}
\end{eqnarray}
and $P_{1}(t)$ is the population in the one-error subspace at time $t$. The rates $r^\pm_j(t)$ quantify the leakage into and out of the codespace per unit time. In the absence of a control Hamiltonian these rates are simply proportional to ${\rm Re}\{\int_0^t d\tau C_j(\tau)\}$ a property of the environmental fluctuations alone. However the control Hamiltonian, in the case of DD and EGP, has the effect of modulating this integral by an oscillating function and hence decreasing its modulus if the rate of oscillation is large enough.

At this point we pause to point out an important difference between error suppression by DD and by EGP. Note that the fact that the DD modulation functions, $m^{\rm DD}_j(t,\tau)$, are real implies that $r_j^+ = r_j^-$ always, regardless of the model of the bath. This highlights a fundamental difference between DD and EGP: EGP imposes a real energy difference between the stabilizer syndrome subspaces and hence bath-induced transition rates between them follow detailed balance (for a bath near thermal equilibrium). This is in contrast with DD that does not impose a real energy gradient and thus all stabilizer syndrome subspaces remain energetically degenerate. Therefore transitions between stabilizer syndrome subspaces do not require energy exchange with the bath and hence associated transitions rates are not Boltzmann weighted. However, note that both techniques, DD and EGP, \emph{suppress} transition rates as a result of modulating the integrands in \erf{eq:r_rates}. The difference between these techniques will become more important when we consider error \textit{correction} in the next section.

\subsubsection{Example: Classical stochastic noise model}
\label{sec:eg_classical_noise}
To illustrate the error suppression consider a classical approximation of the environment (\eg the Kubo-Anderson stochastic model) in which case the correlation function is purely real, and fix it to be exponentially decaying. In this case, $C(t) \propto e^{-\gamma t}$ where $\gamma$ is the inverse correlation time. If in addition the noise amplitude is Gaussian distributed, this describes an Ornstein-Uhlenbeck stochastic process. Consider the case of EGP where the modulation function is sinusoidal, in which case, 
\begin{equation}
r^\pm_j(t) \propto \frac{2 \left(\gamma - \gamma e^{-\gamma t } \cos[\omega_j t]+\omega_j e^{-\gamma t }  \sin[\omega_j t]\right)}{\omega_j^2+\gamma ^2}
\label{eq:r_plusminus}
\end{equation}
where $\omega_j = 2\alpha w_j/\hbar$. Note that this classical model of the bath will not capture relaxation and temperature effects correctly. This is the reason that temperature does not appear in the rates above and that $r_j^+ = r_j^-$. However we consider it here because of the simple form of the resulting correlation function, which in turn allows us to transparently illustrate the mechanisms of error suppression. A more complete calculation with a quantum correlation function that incorporates temperature effects and thus results in unequal upward and downward rates is presented in \ref{sec:full_rate_calc}.

In the expression in \erf{eq:r_plusminus}, the consequences of adding the energy penalty terms are summarized by the presence of the factor $\omega_j$. This term increases with the energy penalty ($\alpha$) and the number of that anti-commute with the error ($w_j$). The term has two effects: (i) its presence in the denominator decreases the overall rate of population leakage, (ii) it increases the oscillation frequency of the sinusoidal functions in the numerator, thus decreasing the magnitude of integrals of $r^\pm_j(t)$ as long as this oscillation frequency is large. Therefore, this calculation explicitly shows how the control Hamiltonian decreases population leakage from the codespace. 

For use in later sections and to connect to previous results on error suppresion \cite{Jordan:2006jb} we also consider the population transfer rates in \erf{eq:r_rates} under the second Markov approximation, which sets the upper limit of the rate integrals to infinity ($t \rightarrow \infty$). Under this further approximation the rates become related to the Fourier (for EGP) or Walsh-Hadamard (for DD) transforms of the bath correlation function. For example, for EGP $r_j^\pm = \frac{2}{\hbar^2} \Re {\sf C}_j( \pm \omega_j)$ under the second Markov approximation, where ${\sf C}_j(\omega) \equiv \int_0^\infty C_j(\tau) e^{i\omega \tau} d\tau$ is the (one-sided) Fourier transform of $C_j(\tau)$. Assuming a harmonic thermal bath, and using symmetries of $C_j(\tau)$, this rate can also be written as  \cite{Ish.Fle-2009a}
\beq
r_j^\pm = \frac{2\mathcal{J}_j(\pm \omega_j)}{\hbar}[\textsf{n}( \pm \omega_j)+1],
\label{eq:secondmarkov_rate}
\eeq 
where $\textsf{n}(\omega) = 1/(e^{\beta\hbar\omega}-1)$ is the average occupation number (according to the Bose-Einstein distribution) and $\mathcal{J}_j(\omega)$ is the spectral density of the bath (with symmetry: $\mathcal{J}(-\omega)=-\mathcal{J}(\omega)$) \footnote{This expression is strictly only valid for $\omega_j=2\alpha w_j/\hbar \neq 0$, otherwise it diverges. The issue is that the second Markov limit, $t\rightarrow \infty$, and the $\omega_j \rightarrow 0$ limit have to be taken carefully when both are required. A careful analysis \cite{Bre.Pet-2002} reveals that when $\omega_j \rightarrow 0$, these rates become $r^{\pm}_j = \frac{2 k_{\rm B} T}{\hbar^2} \frac{\partial \mathcal{J}(\omega)}{\partial \omega} |_{\omega=0}$.}. 

This expression for the rates in the second Markov approximation makes it clear that for large energy penalties the rates are largely determined by the cut-off (or regularization) behavior of the bath spectral density. That is, as the energy penalty, $\alpha$, increases the population transfer rates are proportional the spectral density $\mathcal{J}_j(\omega_j)$ at higher values of $\omega_j$. Realistic spectral densities decay at frequencies above some cut-off, and if the cut-off behavior is Lorentzian like in the Kubo-Andersen model, then the rates will only decay as $\propto \frac{1}{\omega_j^2}$ for large $\omega_j$, while if the spectral density regularization is exponential, the rates decay as $\propto e^{-\omega_j}$ for large $\omega_j$. Therefore knowledge of the high energy behavior of the bath spectral density is critical in assessing the effectiveness of error suppression using EGP or DD in many-qubit systems. This issue was also noted by Jordan \emph{et al.} in Ref. \cite{Jordan:2006jb}. If we require the rates to be suppressed exponentially in $n_l$, the number of logical qubits, then we require that $\omega_j$ (and consequently $\alpha$ or $w_j$) scale exponentially in $n_l$ for a bath with Lorentizian regularization and linearly for a bath with exponential regularization. These are very different requirements, with the former being much more demanding.

\subsection{$H_{\rm AQC} \neq 0$}

The classical rate equation in \erf{eq:pop_eqn} describing subspace population changes is only possible when $H_{\rm AQC}=0$. When this is not the case, the states in a single subspace, \eg the codespace, have different transition rates, as opposed to \erf{eq:pop_eqn} where all states in $\mathbf{P}$ have the same transition rate to $\mathbf{Q_1}$. Therefore a rate equation for subspace populations is no longer possible. Despite this, the effect of the control terms (EGP or DD) is still to suppress transitions between subspaces arising from the environmental coupling. To see this, consider the instantaneous ground state population (in the toggling frame): $P_{\psi_0} = \bra{\psi_0(t)} \tilde \rho(t) \ket{\psi_0(t)}$. The rate of change of this population is given by 
\begin{equation*}
\frac{{\rm d}P_{\psi_0}}{{\rm d}t} = \langle \dot \psi_0(t) | \tilde \rho(t)  \ket{\psi_0(t)} + \bra{\psi_0(t)} \tilde \rho(t) | \dot \psi_0(t) \rangle + \bra{\psi_0(t)} \frac{{\rm d} \tilde \rho(t)}{{\rm d}t} \ket{\psi_0(t)} \nonumber
\end{equation*}
The first two terms represent coherent deformations to the ground state due to adiabatic evolution. These might lead to change in the ground state population through diabatic transitions. But since these cannot be suppressed with our encoding we ignore these, and instead only consider the change in ground state population due to the last term, which induces incoherent transitions. Using \erf{eq:time-local_me} this term evaluates to
\bqa
\fl \frac{{\rm d}P_{\psi_0}}{{\rm d}t} \bigg|_{\rm incoh} = \frac{2}{\hbar^2} \sum_{j=1}^{N_e} \int_0^t d\tau & ~ \Re \bigg\{ C_j(\tau) m_j(t,\tau) \bra{\psi_0(t)} \hat{\Xi}(t,\tau) \tilde{\rho}(t) E_j  \ket{\psi_0(t)} \bigg\} \nn \\
&- \Re \bigg\{ C_j(\tau) m^*_j(t,\tau) \bra{\psi_0(t)} E_j \hat{\Xi}(t,\tau) \tilde{\rho}(t) \ket{\psi_0(t)} \bigg\}
\eqa
Now consider the case where $\tilde{\rho}(t) =\ket{\psi_0(t)}\bra{\psi_0(t)}$. That is, calculate the rate of change in ground state population when the system begins in the ground state. This simplifies to
\bqa
\fl \frac{{\rm d}P_{\psi_0}}{{\rm d}t} \bigg|_{{\rm incoh}, \rho(t)=\ket{\psi_0(t)}\bra{\psi_0(t)}} = - \frac{2}{\hbar^2} \sum_{j=1}^{N_e} \int_0^t d\tau & ~ \Re \bigg\{ C_j(\tau) m^*_j(t,\tau) \bra{\psi_j(t)} e^{-i \bar{H}_{\rm AQC}(t)\tau}\ket{\psi_j(t)} \bigg\} \nn
\eqa
where $\ket{\psi_j(t)} = E_j\ket{\psi_0(t)}$, and we have set the ground state energy to be zero without loss of generality ($\bar{H}_{\rm AQC}(t) \ket{\psi_0(t)}=0$). Furthermore, we have approximated $\hat{\Xi}(t,\tau) \approx \Xi(t,\tau)$ as before. This expression quantifies the rate of population leakage from the ground state as a result of incoherent transitions. The matrix element $\bra{\psi_j(t)} e^{-i \bar{H}_{\rm AQC}(t)\tau}\ket{\psi_j(t)}$ cannot be simplified in general because the error state $\ket{\psi_j(t)}$ is not necessarily an eigenstate of $\bar{H}_{\rm AQC}(t)$ because $[\bar{H}_{\rm AQC}, E_j] \neq 0$. This problem, that error states are not eigenstates of the adiabatic Hamiltonian is a major issue for error correction in AQC, and we shall return to this issue in the next section. However, regardless of the value of this matrix element the above expression confirms that the mechanism of error suppression in the presence of the adiabatic interpolation is the same as when $\bar{H}_{\rm AQC}=0$; \ie the modulation functions add oscillatory components to the dissipation kernels and hence the integrals defining the rate of population leakage can be suppressed as long as this rate of oscillation is large enough (the correlation function will decay quickly after a cut-off frequency for any physical model of the bath and the oscillation frequency, $\omega_j$, should be larger than this cut-off frequency).

\section{Error correction in AQC}
\label{sec:error_corr}
There is no established approach for error correction in AQC. The most obvious approach is to freeze adiabatic evolution at regular intervals, measure the stabilizer generators, and then apply a correction if necessary before recommencing adiabatic evolution. This approach, which we shall refer to as the \textit{freeze-measure-correct} approach, resembles circuit-model error correction but is fraught with practical difficulties. For example, the multi-body measurements necessary for error correction will likley be implemented non-adiabatically and therefore could disturb the ground state population. More importantly, any leakage outside the codespace between error correction cycles becomes uncorrectable due to the issue identified above that error states are not necessarily eigenstates of the adiabatic Hamiltonian. To make this issue more explicitly, we summarize the reasoning presented in Ref. \cite{Young:2013}, where an example evolution was considered: unperturbed evolution of the system until time $\tau$, at which point there is a correctable error $E_j$, proceeded by unperturbed evolution again until an error correction cycle. The optimal error correction operation is the application of $E_j$ again after decoding. Thus the overall evolution is: $\ket{\psi(t)} = E_j \mathcal{U}_{\rm AQC}(t,\tau) E_j \mathcal{U}_{\rm AQC}(\tau,0) \ket{\psi_0)}$. Note that we are only considering evolution in $\mathcal{H}_{\rm sys}$ since this is sufficient and we are ignoring evolution by the control Hamiltonian since it is inconsequential for the following argument. Since the evolution till $\tau$ is unperturbed and adiabatic this is equivalent to: $\ket{\psi(t)} = E_j \mathcal{U}_{\rm AQC}(t,\tau) E_j \ket{\psi_0(\tau)}$ where $\ket{\psi_0(\tau)}$ is the ground state of the adiabatic Hamiltonian at time $\tau$ (and in the codespace). To simplify this further we want to commute the $E_j$ past the AQC unitary. However, the commutation relation between $\mathcal{U}_{\rm AQC}$ and $E_j$ is non-trivial. We first decompose the encoded AQC  Hamiltonian into terms that commute and anti-commute with $E_j$: $\bar{H}_{\rm AQC}(t) = \bar{H}^+_j(t) + \bar{H}^-_j(t)$, with $E_j \bar{H}^{\pm}_j(t) \pm \bar{H}^\pm_j(t) E_j = 0$. With this decomposition, $\bar{H}_{\rm AQC}(t) E_j = E_j (\bar{H}^+_j(t) - \bar{H}^-_j(t))~~\forall t$. Using this to commute the error past the AQC evolution results in: $\ket{\psi(t)} = \overline{\mathcal{U}}_{\rm AQC} \ket{\psi_0(\tau)}$, where
\begin{equation}
\overline{\mathcal{U}}_{\rm AQC} = \exp_+\bigg(-\frac{i}{\hbar}\int_{\tau}^t (\bar H_j^+(s)-\bar H_j^-(s))ds \bigg) \nonumber
\end{equation}
Therefore even after the correction has been applied at time $t$ we do not recover the correct state, $\ket{\psi_0(t)}$. This is in effect because the error state $E_j \ket{\psi_0(\tau)}$ is not an eigenstate of $\bar{H}_{\rm AQC}$ and therefore once a state is promoted into an error subspace $\bar{H}_{\rm AQC}$ coherently mixes it with other states in that error subspace (but not between subspaces), which results in faulty correction. Another way to interpret this problem is to see that the correctable, low weight error $E_j$ is quickly ``dressed" by the adiabatic Hamiltonian into a high weight, uncorrectable error \footnote{The \textit{weight} of a multi-qubit Pauli operator is the number of non-identity terms in the tensor product.}. This is analogous to an error during the implementation of a non-transversal gate in the circuit model. This problem of leaked populations being uncorrectable implies that the error correction cycles in the \textit{freeze-measure-correct} approach have to be extremely frequent. 

In this work we construct an alternative to the \textit{freeze-measure-correct} approach to error correction in AQC. It operates continuous-in-time, does not require the freezing of adiabatic evolution, and relies on cooling. It is known that cooling is analogous to error correction since both are entropy reduction methods; \eg \cite{Bar.War-2000,Sar.Mil-2005, Young:2012wl}. However, the cooling dynamics has to be engineered so that the correct degrees of freedom (erroneous excitations in the case of AQC) are being damped. Cooling of local degrees of freedom is the most experimentally practical and we restrict our attention to such local cooling \footnote{Non-local cooling of \textit{arbitrary} degrees of freedom of a many-body system is an powerful resource that enables efficient error correction and state preparation \cite{Diehl:2008ha, Her.You.etal-2009} but is physically unrealistic.}. In the case of AQC with qubits, local cooling refers to cooling individual qubits independently. In the following we formulate the structure necessary for implementing error correction during AQC with local cooling, and the effective dynamics of encoded AQC with such cooling.

\subsection{Structure required for error correction}
Two specific features are required for error correction by local cooling to be successful. The first is stabilizer encodings and penalties on errors enforced by EGP. DD is insufficient for error correction by cooling because it does not impose real energy penalties. Cooling preferentially biases the system towards low energy states and therefore energy penalties on error subspaces are necessary for the codepsace to be preferentially populated by cooling. Therefore in the following we will assume that the control Hamiltonian implements EGP, and $m_j(t,\tau) = m^{\rm EGP}_j(t,\tau)$ always. 

The second feature required for error correction by cooling is related to the above observation that leakage from the codespace is irrecoverable due to coherent mixing of states in the errror subspaces by the adiabatic Hamiltonian. In order to circumvent this problem we must modify the construction of the encoded adiabatic Hamiltonian $\bar{H}_{\rm AQC}$. To understand how to do this modification it is useful to present another interpretation of why the problem exists. The essence of the problem is that \emph{local perturbations (by single qubit Pauli errors) of the many-body ground state of $\bar{H}_{\rm AQC}+H_{\rm C}$ quickly become non-local excitations of the many-body system due to the couplings induced by the Hamiltonian}. Therefore, subsequent cooling of local degrees of freedom (single qubits) cannot destroy this delocalized excitation. This is not an issue in quantum memories created from degenerate ground states of stabilizer Hamiltonians (\eg the abelian toric code) because in these cases local perturbations create excitations that remain localized and therefore can be subsequently quenched by local cooling \cite{Young:2012wl}. Given this perspective it is clear that a way to fix this problem is to modify the logical AQC Hamiltonian $\bar{H}_{\rm AQC}$ so that local perturbations create localized excitations, \ie the correctable error states remain eigenstates of the encoded Hamiltonian. This can in fact always be done because of the freedom in choice of logical operators in stabilizer codes; a logical operator can be multiplied by any linear combination of stabilizer generators to create an equivalent logical operator. We will refer to these modified logical Hamiltonians as \textit{protected Hamiltonians} and notate them as $\bar{H}^\lambda_{\rm AQC}$.

\subsection{Protected Hamiltonians for AQC}

In order to specify the algebraic properties of the protected Hamiltonians we must first introduce some notation. The system bath interaction, $\sum_{j=1}^{N_e} E_j\otimes B_j$, defines the elementary errors in the system as $E_j$. We specialize to the case where the physical system-environment interaction contains single qubit error terms, \ie $E_j$ in \erf{eq:aqchambar} acts non-trivially only on one qubit and $E_j^2 = \mathbf{1}$. This is a physically realistic assumption since the system-environment interaction is likely low weight. However, the code we use to encode the system could correct more than one error, \eg could have distance $d>3$. In this case, the system is recoverable even after multiple errors. 

\begin{figure}
\centering
\includegraphics[scale=0.7]{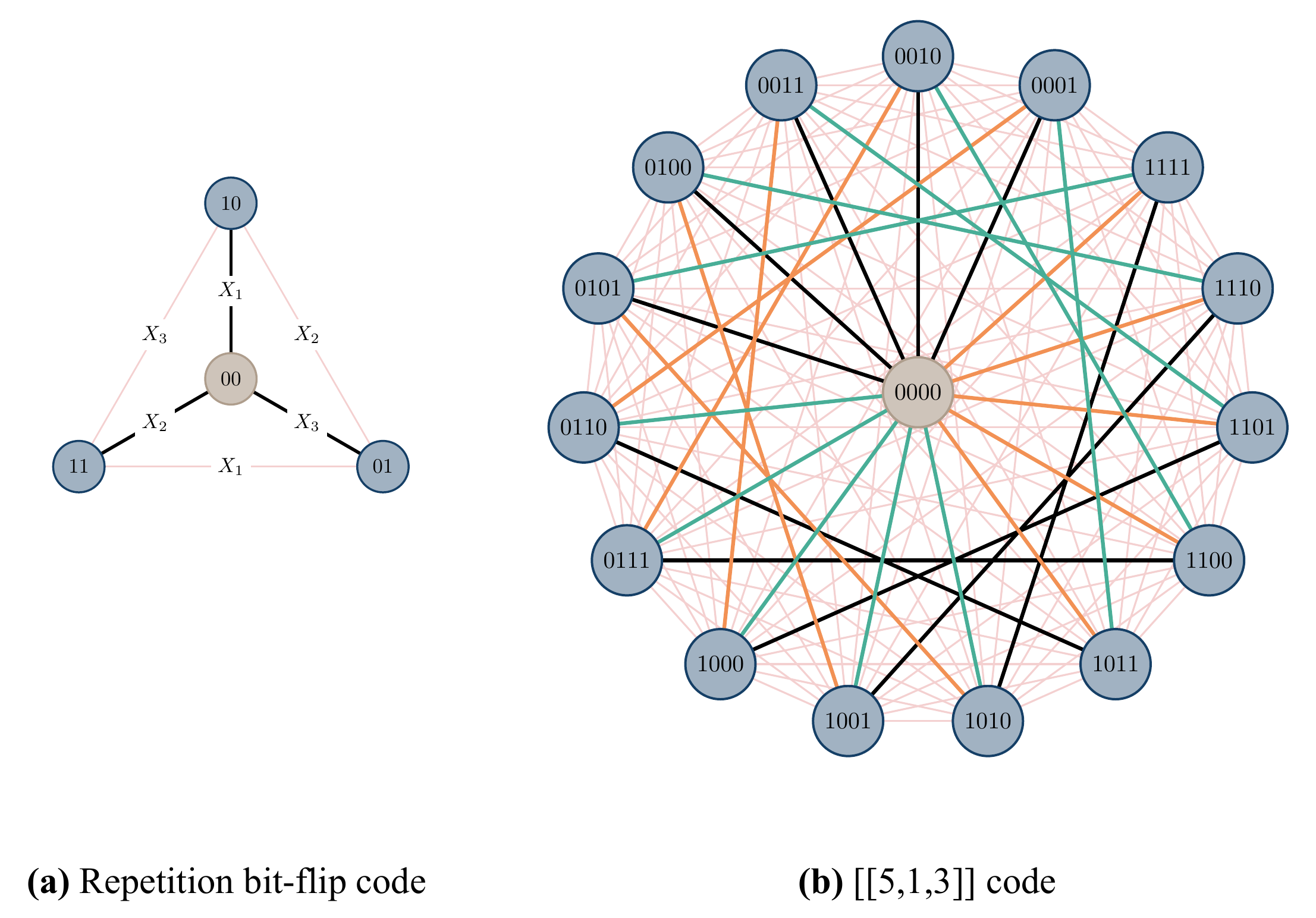}
\caption{\small \label{fig:ecc_spaces} The structure of encoded Hilbert space and transitions between syndrome subspaces induced by incoherent single qubit (elementary) errors.  An $[[n,k,d]]$ quantum code encodes $k$ logical qubits in $n$ physical qubits and corrects at least $\lfloor (d-1)/2 \rfloor$ Pauli errors. The two examples shown here are \textbf{(a)} the repetition bit-flip code with stabilizer generators $ZZI$ and $IZZ$ (not a full quantum code since phase errors are not corrected), and \textbf{(b)} the [[5,1,3]] code with stabilizer generators $IXZZX, XIXZZ, ZXIXZ$ and $ZZXIX$. In each example the circles represent $2^{k}$ dimensional syndrome subspaces of the encoded Hilbert space. These subspaces are labelled by a syndrome pattern $\bnu = \nu_1 \nu_2,...\nu_{n-k}$, a vector whose binary entries denote whether each one of the $n-k$ stabilizer generators commutes ($\nu_i=0$) or anti-commutes ($\nu_i=1$) with the error $E_\bnu$ that takes a state in the codespace to that syndrome subspace. $\bnu=\boldsymbol{0}$ represents the codespace. Both these examples are perfect, non-degenerate codes so each syndrome subspace corresponds to a unique correctable error. Black, green and orange lines correspond to the transitions between subspaces induced by correctable single qubit $\sigma_x$, $\sigma_z$, and $\sigma_y$ errors, respectively. The red lines corresponds to transitions induced by uncorrectable errors. We have labeled the black edges with the corresponding transition-inducing errors for the bit-flip code but have omitted these labels for clarity for the larger $[[5,1,3]]$ code. The rate equation dynamics derived in the main text describes a Markov random walk between these subspaces because the elementary errors $E_j$ move states incoherently between the syndrome subspaces. Beginning in the codespace, the overall state is guaranteed to be correctable as long as no red line is crossed during the random walk. }
\end{figure}

A general sequence of elementary errors, $\prod_{i=1}^l E_{j_i}$ with $1\leq j_i \leq N_e$ can build up to a correctable or uncorrectable error. By the error correction conditions \cite{mikeandike}, each correctable error is identified by a unique anti-commutation pattern with the stabilizer generators of the code, the \textit{syndrome pattern}. Motivated by this we denote the Pauli operator associated to a sequence of elementary errors as $E_{\bnu}$, where the label $\bnu$ is a binary vector indicating which stabilizer \textit{generators} anti-commute with the error. Explicitly, $\bnu = (\nu_1, \nu_2, ... \nu_g)$, where $g$ is the number of generators in the code (an $[[n,k,d]]$ stabilizer code has $g=n-k$ generators), and $E_\bnu S_m = (-1)^{\nu_m}S_m E_\bnu$ where $S_m$ is a stabilizer generator \footnote{In the following, a ``no error" is considered to be in the set of correctable errors with $E_{\bnu = \boldsymbol{0}} = \mathbf{1}$.}. Each elementary error $E_j$, which are all assumed to be correctable by the employed code, has a syndrome pattern that we denote $\bnu(j)$. Therefore the elementary errors could be alternatively written as $E_{\bnu(j)}$. Finally, we also define projectors onto the $2^k$-dimensional syndrome subspaces as $\mathbf{Q}_\bnu = E_\bnu  \mathbf{P} E_\bnu$. Note that $\sum_{\bnu} \mathbf{Q}_\bnu= \mathbf{1}$ is a resolution of identity, where $\mathbf{Q}_{\boldsymbol{0}} = \mathbf{P}$ is included in the sum. See Fig. \ref{fig:ecc_spaces} for a graphical representation of syndrome subspaces and some of the above definitions in encoded Hilbert space. 

Note that for perfect codes \cite{Got-1997} the number of syndrome patterns, $2^{n-k}$, exactly equals the number of correctable errors. Whereas, for imperfect codes the number of syndrome patterns $\bnu$ could be larger than the set of correctable errors. Therefore one must be careful when tracking error sequences using the syndrome labeling. While each correctable error sequence produces a unique syndrome pattern, an uncorrectable error sequence can produce a syndrome pattern that is the same as, or could be different from (for imperfect codes) the syndrome pattern of a correctable error sequence. Starting from the codespace, the syndrome pattern uniquely labels an error sequence as long as it is a correctable error, but as we build up more and more errors we must be careful to track when a composite error becomes uncorrectable. For example, given two correctable errors $E_\bnu$ and $E_{\boldsymbol{\mu}}$, the concatenated error $E_{\bnu}E_{\boldsymbol{\mu}} = E_{\bnu \oplus \boldsymbol{\mu}}$ ($\oplus$ denotes binary addition) could be another correctable error or an uncorrectable error. The syndrome label $\bnu \oplus \boldsymbol{\mu}$ by itself does not tell us which one it is in all cases. Finally, we note that while the distance of the quantum code is a useful proxy for identifying correctable states, it is not always sufficient. That is, for a non-degenerate code the correctable errors are the ones with weight $\leq \lfloor (d-1)/2 \rfloor$, where $d$ is the distance of the code. However, degenerate codes can correct errors that have weight greater than $\lfloor (d-1)/2 \rfloor$. Therefore to be general and capture degenerate as well as imperfect quantum codes, we will simply refer to an error sequence $\prod_{i=1}^l E_{j_i}$ with $1\leq j_i \leq N_e$ as being a \textit{correctable} or \textit{uncorrectable} error. Either way, it can be associated with a syndrome pattern $\bnu$, and when there is no ambiguity we will label it as $E_\bnu$ \footnote{For degenerate codes it is more correct to refer to $E_\bnu$ as the ``correction operation" since $\bnu$ labels a syndrome subspace and multiple correctable errors can map to the same syndrome space for a degenerate code. But for simplicity we refer to $E_\bnu$ as an error with this subtlety implied for degenerate codes.}.

Moreover, given this notation we can label a (time dependent) complete basis in the Hilbert space of the encoded system as $\ket{\epsilon_n(t), \bnu}$ with $\bar{H}^\lambda_{\rm AQC}(t) \ket{\epsilon_n(t), \boldsymbol{0}} = \epsilon_n(t) \ket{\epsilon_n(t), \boldsymbol{0}}$; \ie the quantum numbers $\epsilon_n$ label the eigenvalues of $\bar{H}^\lambda_{\rm AQC}(t)$ for a state in the codespace. Furthermore, $\sum_{n, \bnu} \ket{\epsilon_n(t), \bnu}\bra{\epsilon_n(t), \bnu} = \mathbf{1}$, where the sum over $\bnu$ is over all $2^{n-k}$ (correctable and uncorrectable) syndrome patterns.

Given this notation, the fundamental property of a protected implementation of the logical Hamiltonian is that all correctable errors take eigenstates in the codespace to eigenstates in error subspaces. That is, for a correctable error $E_\bnu$,  
\beq
\bar{H}^\lambda_{\rm AQC}(t) E_\bnu\ket{\epsilon_n(t), \boldsymbol{0}} = \epsilon_{n,\bnu}(t) E_\bnu \ket{\epsilon_n(t), \boldsymbol{0}} = \epsilon_{n,\bnu}(t) \ket{\epsilon_n(t), \boldsymbol{\bnu}}
\eeq 
This condition stipulates that the erred (but correctable) state is an eigenstate of the protected Hamiltonian. In principle $\epsilon_{n,\bnu}$ can have arbitrary dependence on $n$ and $\bnu$, but for practical constructions based on exploiting the stabilizer structure (see Ref. \cite{Young:2013}) this energy factorizes into $\epsilon_{n,\bnu}(t) = \epsilon_n(t) \lambda_\bnu$. The first factor is the same as the energy in the codespace while $\lambda_\bnu \in \mathbb{R}$ ($\lambda_\bnu \neq 0$) is a deformation factor that modifies the energy of the corresponding state in the error space. From demanding this property we get the fundamental property of protected logical Hamiltonians, that for all correctable errors:
\beq
\bar{H}^\lambda_{\rm AQC}(t) E_\bnu \mathbf{P} - \lambda_\bnu E_\bnu \bar{H}^\lambda_{\rm AQC}(t) \mathbf{P} = 0 ~~~~ \forall t, ~ E_\bnu ~{\rm correctable}
\label{eq:prot_ham_prop}
\eeq
This can be viewed as a deformed commutator between the logical Hamiltonian and the correctable errors (but only when operating on the codespace). Note that although $\bar{H}^\lambda_{\rm AQC}$ is time-dependent $\lambda_\bnu$ is time independent; its only dependency is on the error syndrome.

Although \erf{eq:prot_ham_prop} is a property that we are demanding from protected Hamiltonians, we can also show that it is always possible to construct a logical AQC Hamiltonian that satisfies this property. A constructive algorithm is given Ref. \cite{Young:2013} and results in protected Hamiltonians of the following form
\beq
\bar{H}^{\lambda}_{\rm AQC}(t) = \sum_{E_\bnu {\rm ~correctable}} \lambda_\bnu E_{\bnu} \bar{H}_{\rm AQC}(t) \mathbf{P} E_{\bnu}
\eeq
where $\bar{H}_{\rm AQC}$ is a conventional encoded AQC Hamiltonian using arbitrary logical operators of the code (note that $\mathbf{P} \bar{H}_{\rm AQC} \mathbf{P} = \bar{H}_{\rm AQC} \mathbf{P}$, since the logical Hamitlonian does not connect different stabilizer subspaces, and hence $\bar{H}^{\lambda}_{\rm AQC}$ is Hermitian). Although this protected Hamiltonian is typically a very high weight operator, it is possible to decrease its weight by utilizing the structure of the code \cite{Young:2013}. However, the weight of the protected Hamiltonian can never be decreased below $d$ since it must contain logical operators of the code. A distinguished protected Hamiltonian, which has the property that $\lambda_\bnu =1, ~\forall \bnu$ will be important in the analysis below and we refer to it as the \textit{canonical protected Hamiltonian}, and notate it by $\bar{H}^p_{\rm AQC}$. 

\subsection{Dynamics under protected AQC Hamiltonians}
Before introducing the cooling necessary for error correction we will first derive equations describing non-equilibrium dynamics under a protected Hamiltonian implementation of AQC. The dynamics of the encoded AQC evolution is best described as dynamics of populations in the syndrome subspaces described above. Defining, $P_{\bnu} = \tr(\mathbf{Q}_{\bnu} \tilde{\rho}(t))$ as the population of the syndrome subspace labelled by $\bnu$, \erf{eq:time-local_me} can be used to derive the following evolution:

\bqa
\frac{{\rm d}P_{\bnu}(t)}{{\rm d}t} =  \frac{2}{\hbar^2}\sum_{j=1}^{N_e} \int_0^t d\tau  &\Re \bigg\{  
	 C_j(\tau) \tr \bigg[ \mathbf{Q}_{\bnu} \tilde{\Xi}_j(t,\tau) \tilde{\rho}(t) \tilde{E}_j(t) \mathbf{Q}_{\bnu} \bigg]  \bigg\}  \nn \\
	 &- \Re \bigg\{ C_j(\tau) \tr \bigg[ \mathbf{Q}_{\bnu} \tilde{E}_j(t) \tilde{\Xi}_j(t,\tau) \tilde{\rho}(t) \mathbf{Q}_{\bnu} \bigg] \bigg\}
\label{eq:p_nu_eq}
\eqa
As above we want to use the properties of the control and logical Hamiltonians to simplify the traces in this equation. To begin, we exploit the expression for the elementary error operators in the toggling frame given by \erf{eq:ejegp}. Then, using the approximation $\hat{\Xi}(t,\tau) \approx \Xi(t,\tau) \equiv e^{-\frac{i}{\hbar}\tau \bar H^\lambda_{\rm AQC}(t)} E_j~e^{\frac{i}{\hbar}\tau \bar H^\lambda_{\rm AQC}(t)}$, the first trace becomes:
\begin{eqnarray}
e^{\frac{i}{\hbar}2 \alpha \tau \varpi(j,\bnu)} \tr \bigg[ \mathbf{Q}_{\bnu} e^{-\frac{i}{\hbar}\bar{H}^\lambda_{\rm AQC}(t)\tau} E_j e^{\frac{i}{\hbar}\bar{H}^\lambda_{\rm AQC}(t)\tau} \tilde{\rho}(t) E_j \mathbf{Q}_{\bnu} \bigg]
\end{eqnarray}
where 
\begin{eqnarray}
\varpi(j,\bnu) = \sum_{m ~{\rm s.t.} \{S_m,E_j\}=0} (-1)^{\nu_m} ~~=~ \bnu(j) \cdot \left[\bnu(j)\oplus \bnu \right] - \bnu(j) \cdot \bnu
\label{eq:varpi}
\end{eqnarray}
This factor is analogous to the factor $w_j$ in section \ref{sec:error_supp}, however in this case this frequency depends on the entire error sequence and not just $E_j$. 

To simplify this trace further we first utilize the fundamental property of the protected Hamiltonian, \erf{eq:prot_ham_prop}, to establish a similar relation for elementary errors applied to states in error subspaces:
\begin{eqnarray}
\bar{H}^\lambda_{\rm AQC}(t) E_j \mathbf{Q}_{\bnu} - \frac{\lambda_{\bnu(j)\oplus\bnu}}{\lambda_\bnu} E_j \bar{H}^\lambda_{\rm AQC}(t) \mathbf{Q}_{\bnu} = 0 ~~~~ \forall t, j 
\label{eq:prot_ham_prop_Qnu}
\end{eqnarray}
which holds as long as the concatenated error $E_j E_\bnu = E_{\bnu(j)\oplus \bnu}$ is a correctable error. Here $\lambda_{\bnu(j) \oplus \bnu}$ is the deformation factor for the concatenated error; \ie $\bar{H}^\lambda_{\rm AQC}(t) E_{\bnu(j) \oplus \bnu} \mathbf{P} - \lambda_{\bnu(j)\oplus\bnu}(t) E_{\bnu(j) \oplus \bnu} \bar{H}^\lambda_{\rm AQC}\mathbf{P} = 0$. 

This identity simplifies the first trace in \erf{eq:p_nu_eq}, as long as $E_j E_{\bnu}$ is also a correctable error, to
\begin{eqnarray}
& \sum_{n} e^{\frac{i}{\hbar}\tau \left[2\alpha \varpi(j,\bnu) + \epsilon_n(t) (\lambda_{\bnu} - \lambda_{\bnu(j)\oplus \bnu})\right]} \bigg\langle\epsilon_n(t), \bnu(j)\oplus \bnu  \bigg| \tilde{\rho}(t) \bigg| \epsilon_n(t), \bnu(j)\oplus \bnu \bigg\rangle
\label{eq:tr1}
\end{eqnarray}
Similarly, using the above properties, we can simplify the second trace in \erf{eq:p_nu_eq} to:
\begin{eqnarray}
 \tr &\bigg[ \mathbf{Q}_{\bnu} \tilde{E}_j(t) \tilde{\Xi}_j(t,\tau) \tilde{\rho}(t) \mathbf{Q}_{\bnu} \bigg] \nn \\
 &\approx \sum_n e^{-\frac{i}{\hbar}\tau \left[2\alpha \varpi(j,\bnu) + \epsilon_n(t) (\lambda_{\bnu} - \lambda_{\bnu(j)\oplus \bnu})\right]}  \bigg\langle \epsilon_n(t), \bnu \bigg| \tilde{\rho}(t) \bigg|\epsilon_n(t), \bnu \bigg\rangle
 \label{eq:tr2}
\end{eqnarray}
when $E_{\bnu}$ is a correctable error. 

\erf{eq:tr1} is proportional to the gain in population in syndrome subspace $\bnu$ as a result of transitions from syndrome subspace $\bnu(j)\oplus \bnu$. Similarly, \erf{eq:tr2} is proportional to the total loss in population in syndrome subspace $\bnu$ to the neighboring subspace $\bnu(j)\oplus \bnu$. The exponential factor $\pm \left[2\alpha \varpi(j,\bnu) + \epsilon_n(t) (\lambda_{\bnu} - \lambda_{\bnu(j)\oplus \bnu})\right]$ represents the gain or loss in energy as a result of the transition. 
There are two contributions to this energy. The first comes from the energy difference between the syndrome subspaces projected onto by $\mathbf{Q}_\bnu$ and $\mathbf{Q}_{\bnu+j}$. This energy difference is enforced by the EGP Hamiltonian and is equal to $2\alpha \varpi(j,\bnu)$. The second component to the overall energy cost comes from the energy difference caused by the deformation of the energy landscape by the protected logical Hamiltonian, and is given by $\epsilon_n(t) (\lambda_{\bnu} - \lambda_{\bnu(j)\oplus \bnu})$.

These traces almost describe the net population of syndrome subspaces, however the issue is that each state in a syndrome subspace has a different rate of transition to neighboring syndrome subspaces (\ie the exponent in \erf{eq:tr2} is $n$ dependent). We can avoid this scenario if we utilize the canonical protected Hamiltonian, in which case $\lambda_\bnu=1 ~ \forall$ correctable $\bnu$, and the two traces above become:
\bqa
 \tr \bigg[ \mathbf{Q}_{\bnu} \tilde{\Xi}_j(t,\tau) \tilde{\rho}(t) \tilde{E}_j(t) \mathbf{Q}_{\bnu} \bigg] &= e^{\frac{i}{\hbar}\tau \left[2\alpha \varpi(j,\bnu) \right]} P_{\bnu(j)\oplus \bnu} \nn \\
 \tr \bigg[ \mathbf{Q}_{\bnu} \tilde{E}_j(t) \tilde{\Xi}_j(t,\tau) \tilde{\rho}(t) \mathbf{Q}_{\bnu} \bigg] &= e^{-\frac{i}{\hbar}\tau \left[2\alpha \varpi(j,\bnu) \right]} P_{\bnu}  
\eqa
By putting these simplified traces together, the evolution of the syndrome subspace populations, for syndromes that represent correctable errors, follows a classical master/rate equation:
\begin{eqnarray}
\frac{{\rm d}P_{\bnu}(t)}{{\rm d}t} \approx   \sum_{j~{\rm s.t.} \atop E_jE_{\bnu}~{\rm correctable}} R^+_{\bnu,j}(t) P_{\bnu(j)\oplus \bnu}(t) - \sum_{j~{\rm s.t.} \atop E_jE_{\bnu}~{\rm correctable}} R^-_{\bnu,j}(t) P_{\bnu}(t)  
\label{eq:prot_rate_eqn}
\end{eqnarray}
for $\bnu$ correctable, with time-dependent rates
\begin{eqnarray}
R^+_{\bnu,j}(t) &= \frac{2}{\hbar^2}\Re \left\{ \int_0^t d\tau C_j(\tau) e^{\frac{i}{\hbar}\tau\left[2\alpha \varpi(j,\bnu)\right]} \right\} \label{eq:r+} \\
R^-_{\bnu,j}(t) &= \frac{2}{\hbar^2}\Re \left\{ \int_0^t d\tau C_j(\tau) e^{-\frac{i}{\hbar}\tau\left[2\alpha \varpi(j,\bnu)\right]} \right\} \label{eq:r-_prelim}
\end{eqnarray}
Therefore, by utilizing the localizing properties of the canonical protected Hamiltonian we can see that the master equation \erf{eq:p_nu_eq} represents dynamics between syndrome subspace populations. The dynamics resembles a Markov chain random walk with each state in the chain being represented by a syndrome pattern label. 

However, until now we have only accounted for transitions between subspaces caused by correctable errors, \ie transitions along the non-red edges in Fig. \ref{fig:ecc_spaces}. We cannot track the system accurately once it crosses a red edge in Fig. \ref{fig:ecc_spaces} because the syndrome subspaces no longer necessarily uniquely identify states that are uncorrectable. However, we can represent dynamics across this correctable-uncorrectable boundary as a leakage from the correctable subspace of states. To do so, we can sum over all rates of departure from the correctable subspace, which requires evaluation of $ \tr \bigg[ \mathbf{Q}_{\bnu} \tilde{E}_j(t) \tilde{\Xi}_j(t,\tau) \tilde{\rho}(t) \mathbf{Q}_{\bnu} \bigg]$ when $E_j E_\bnu$ is an uncorrectable error and $E_\bnu$ is a correctable error. To evaluate this trace we derive the property:
\bqa
\fl \bar{H}^p_{\rm AQC}(t) E_j \mathbf{Q}_\bnu &=& \sum_{E_\bmu {\rm ~correctable}} E_{\bmu} \bar{H}_{\rm AQC}(t)\mathbf{P} E_{\bmu} E_j E_{\bnu}  \mathbf{P} E_{\bnu} \nn \\
&=& \left\{
\begin{array}{l l}
E_j \bar{H}_\bnu^*(t) \mathbf{Q}_\bnu & \textrm{if $\bnu(j)\oplus \bnu$ is a correctable syndrome pattern} \\
0 &\textrm{if $\bnu(j)\oplus \bnu$ is an uncorrectable syndrome pattern}
\end{array} \right.
\label{eq:uncorr_condns}
\eqa
where the second option is only possible if the code employed is an imperfect code. The second equality follows from considering $\mathbf{P} E_{\bmu} E_j E_{\bnu}\mathbf{P}$; if $\bnu(j)\oplus \bnu$ is an uncorrectable syndrome pattern (which can only be the case if the code is imperfect) then it is distinct from any $\bmu$ in the sum and therefore $\mathbf{P} E_{\bmu} E_j E_{\bnu}\mathbf{P}=0$. On the other hand, if $\bnu(j)\oplus \bnu$ is a correctable syndrome pattern there exists a $\bmu = \bnu(j)\oplus \bnu$ in the sum, and therefore $\mathbf{P} E_{\bmu} E_j E_{\bnu}\mathbf{P} = \delta_{\bmu, \bnu(j)\oplus \bnu}\mathbf{P}$. In this case then, 
$
\bar{H}^p_{\rm AQC}(t) E_j \mathbf{Q}_\bnu = E_{\bnu(j)\oplus \bnu} \bar{H}_{\rm AQC}(t) \mathbf{P} E_{\bnu} = E_j E_\bnu \bar{H}_{\rm AQC}(t) E_\bnu \mathbf{Q}_\bnu = E_j \bar{H}^*_{\bnu}(t)\mathbf{Q}_\bnu
$,
where we have defined 
\beq
\bar{H}^*_{\bnu}(t) \equiv E_\bnu \bar{H}_{\rm AQC}(t) E_\bnu = \bar{H}^+_{{\rm AQC},\bnu}(t) - \bar{H}^-_{{\rm AQC},\bnu}(t)
\eeq 
with $\bar{H}^\pm_{{\rm AQC},\bnu}(t)$ being the terms in $\bar{H}_{\rm AQC}(t)$ that commute/anti-commute with $E_\bnu$. $\bar{H}^*_\bnu$ is still a logical Hamiltonian in the sense that it only contains logical operators, but it is not the same as $\bar{H}^p_{\rm AQC}$. It encapsulates the logical error incurred as a result of concatenating $E_j$ and $E_\bnu$.

Focusing on the first case in \erf{eq:uncorr_condns} for now, where $\bnu(j)\oplus \bnu$ is a correctable syndrome pattern, we have:
\bqa
\fl \tr \bigg[ \mathbf{Q}_{\bnu} \tilde{E}_j(t) \tilde{\Xi}_j(t,\tau) \tilde{\rho}(t) \mathbf{Q}_{\bnu} \bigg] &\approx e^{-\frac{i}{\hbar}\tau [2\alpha \varpi(j,\bnu)]} \tr \bigg[\mathbf{Q}_{\bnu} E_j e^{-\frac{i}{\hbar} \bar{H}^p_{\rm AQC}(t)\tau} E_j e^{\frac{i}{\hbar}\bar{H}^p_{\rm AQC}(t)\tau } \tilde{\rho}(t) \mathbf{Q}_{\bnu} \bigg] \nn \\
&= e^{-\frac{i}{\hbar}\tau [2\alpha \varpi(j,\bnu)]} \tr \bigg[\mathbf{Q}_{\bnu} e^{-\frac{i}{\hbar} \bar{H}^*(t)\tau} e^{\frac{i}{\hbar}\bar{H}^p_{\rm AQC}(t)\tau } \tilde{\rho}(t) \mathbf{Q}_{\bnu} \bigg] \nn \\
&=   \sum_{mn} e^{-\frac{i}{\hbar}\tau [2\alpha \varpi(j,\bnu) -\epsilon_m(t)]} \bra{\epsilon_n(t) , \bnu} e^{-\frac{i}{\hbar} \bar{H}^*(t)\tau}  \ket{\epsilon_m(t), \bnu}  \cdot \nn \\
& ~~~~~~~~~~~~~~~~~~~~~~~~~~~~~~~~~~~~~~~~~~~ \bra{\epsilon_m(t), \bnu} \tilde{\rho}(t) \ket{\epsilon_n(t), \bnu} 
\label{eq:corr-uncorr}
\eqa
where in the second line we have used \erf{eq:uncorr_condns} and in the third we have expanded the projector as $\mathbf{Q}_\bnu = \sum_n \ket{\epsilon_n(t), \bnu}\bra{\epsilon_n(t), \bnu}$,  and used the fact that both $\bar{H}^p_{\rm AQC}$ and $\bar{H}^*$ only contain logical operators. The matrix element $ \bra{\epsilon_n(t) , \bnu} e^{-\frac{i}{\hbar} \bar{H}^*(t)\tau}  \ket{\epsilon_m(t), \bnu} $ is in general going to be non-zero and carry a dependence on $n,m$ and $\tau$, and to simplify this further we need to make some approximations. 

The first approximation we make is $e^{-\frac{i}{\hbar}\tau [2\alpha \varpi(j,\bnu) -\epsilon_m(t)]} \approx e^{-\frac{i}{\hbar}\tau [2\alpha \varpi(j,\bnu) -\bar{\epsilon}(t)]}$ where $\bar{\epsilon}(t) = \frac{1}{2^k}\sum_{m=1}^{2^k} \epsilon_m(t)$ is the mean energy of the states in a syndrome subspace \footnote{Note that since we are using the canonical protected Hamiltonian, the deformation factor $\lambda_\bnu=1$ and hence there is no $\bnu$ dependence on this average energy.}. This is a reasonable approximation because in the regime of good error correction/suppression the EGP energy penalty will be greater than the energy spread within a syndrome subspace, \ie $\alpha \gg \epsilon_m(t) ~\forall m,t$, and hence we can replace the $m$ dependent value with the average since the oscillation frequency will be mostly determined by the first term $2\alpha \varpi(j,\bnu)$.

The second approximation is that the matrix element $\bra{\epsilon_m(t), \bnu} \tilde{\rho}(t) \ket{\epsilon_n(t), \bnu}$ is only non-zero when $n=m$ since we take the adiabatic interpolation to be slow enough to avoid diabatic errors. Hence there is no coherence between logical eigenstates in a syndrome space. Finally, we assume that for small $\tau$ (because of the decaying correlation function $C_j(\tau)$, the values of the above trace for small values of $\tau$ are most important), $\bra{\epsilon_m(t), \bnu} e^{-\frac{i}{\hbar} \bar{H}^*(t)\tau} \ket{\epsilon_m(t), \bnu} \approx  1$. That is, the rotation by the error unitary will be negligible for the $\tau$ values of relevance. This is obviously a crude approximation (but one that improves in quality as the correlation time of the environment decreases), and we comment on ways to refine it below.

Using these approximations, we can estimate the above trace, when $E_j E_\bnu$ is an \emph{uncorrectable} error as
\beq
 \tr \bigg[ \mathbf{Q}_{\bnu} \tilde{E}_j(t) \tilde{\Xi}_j(t,\tau) \tilde{\rho}(t) \mathbf{Q}_{\bnu} \bigg] \approx e^{-\frac{i}{\hbar}\tau \left[2\alpha \varpi(j,\bnu) - \bar{\epsilon} \right]} P_{\bnu}  
\eeq
It can be shown using the same approximations that this same expression holds if the second case of \erf{eq:uncorr_condns} is true. Thus we now have a complete rate model for the error dynamics:
\begin{eqnarray}
\frac{{\rm d}P_{\bnu}(t)}{{\rm d}t} \approx   \sum_{j~{\rm s.t.} \atop E_jE_{\bnu}~{\rm correctable}} R^+_{\bnu,j}(t) P_{\bnu(j)\oplus \bnu}(t) - \sum_{j} R^-_{\bnu,j}(t) P_{\bnu}(t)  
\label{eq:prot_rate_eqn2}
\end{eqnarray}
for $\bnu$ correctable, with $R_{\bnu,j}^+(t)$ as above and 
\beq
\fl R_{\bnu,j}^-(t) \equiv \left\{
\begin{array}{l l}
\frac{2}{\hbar^2}\Re \left\{ \int_0^t d\tau C_j(\tau) e^{-\frac{i}{\hbar}\tau\left[2\alpha \varpi(j,\bnu)\right]} \right\} & \textrm{if $E_j E_\bnu$ is a correctable error} \\
\frac{2}{\hbar^2}\Re \left\{ \int_0^t d\tau C_j(\tau) e^{-\frac{i}{\hbar}\tau\left[2\alpha \varpi(j,\bnu) - \bar{\epsilon}(t)\right]} \right\} & \textrm{if $E_j E_\bnu$ is an uncorrectable error} 
\end{array} \right.
\label{eq:r-}
\eeq
It is important that the first sum in \erf{eq:prot_rate_eqn2} is over all elementary errors such that $E_jE_\bnu$ is correctable, while the second sum is over \textit{all} elementary errors. This means that we are tracking only population within the correctable subspace of states -- population that leaks out into the uncorrectable subspace is not accounted for. Consequently, the net population in the correctable subspace, $P_{\rm corr} \equiv \sum_{\bnu~{\rm correctable}} P_\bnu$, will decrease over time. This correctable population lower bounds the probability that the AQC computation has not failed (due to environment-induced failure modes). This is especially useful for examining the effect of error correction by local cooling which attempts to keep the net error weight small and enhance this success probability. 

We note that most of the approximations used above to simplify the expressions for rates across the correctable-uncorrectable boundary can be avoided at the expense of tracking the leakage rate of each state across this boundary. That is, only by assuming the absence of diabatic errors, we can write \erf{eq:corr-uncorr} as:
\bqa
\fl \tr \bigg[ \mathbf{Q}_{\bnu} \tilde{E}_j(t) \tilde{\Xi}_j(t,\tau) \tilde{\rho}(t) \mathbf{Q}_{\bnu} \bigg] &\approx  \sum_{m} e^{-\frac{i}{\hbar}\tau [2\alpha \varpi(j,\bnu) -\epsilon_m(t)]} h_m(\tau, t) \bra{\epsilon_m(t), \bnu} \tilde{\rho}(t) \ket{\epsilon_m(t), \bnu} \nn
\eqa
where $h_m(\tau, t) \equiv  \bra{\epsilon_m(t) , \bnu} e^{-\frac{i}{\hbar} \bar{H}^*(t)\tau}  \ket{\epsilon_m(t), \bnu}$. Thus each state in the $\bnu$ syndrome subspace has a different rate of leakage to the uncorrectable $\bnu(j)\oplus \bnu$ subspace. We simply have to keep track of these different rates across this boundary and sum them up separately in order to improve the model. We do not explicitly do that here since the rate equation above is accurate enough for our purposes.

To summarize this section, we have been able to derive a description of encoded AQC evolution that completely decouples the adiabatic evolution and the environmentally induced error evolution. This was possible because of the structure of the encoded Hilbert space and properties of stabilizer codes that ensure that logical evolution and evolution due to errors are orthogonal. We note that the above Markov chain random walk description of error-induced evolution requires several ingredients: (i) an almost Markov description of the environment, (ii) the imposition of energy penalties for erroneous states by an EGP control Hamiltonian, and (iii) a protected implementation of the logical Hamiltonian that localizes correctable errors. 

We draw attention to previous works on continuous-time error correction that described the optimal tracking of errors as a Markov chain random walk on syndrome subspaces \cite{Han.Mab-2005a, Cha.Lan.etal-2008, Mabuchi:2009vn}. However, in contrast to the present work, these formulations were for quantum memories and hence did not consider logical Hamiltonians, and further, did not explicitly consider physical system-environment models for the error dynamics.

\subsubsection{Example}
\begin{figure}
\centering
\includegraphics[scale=0.6]{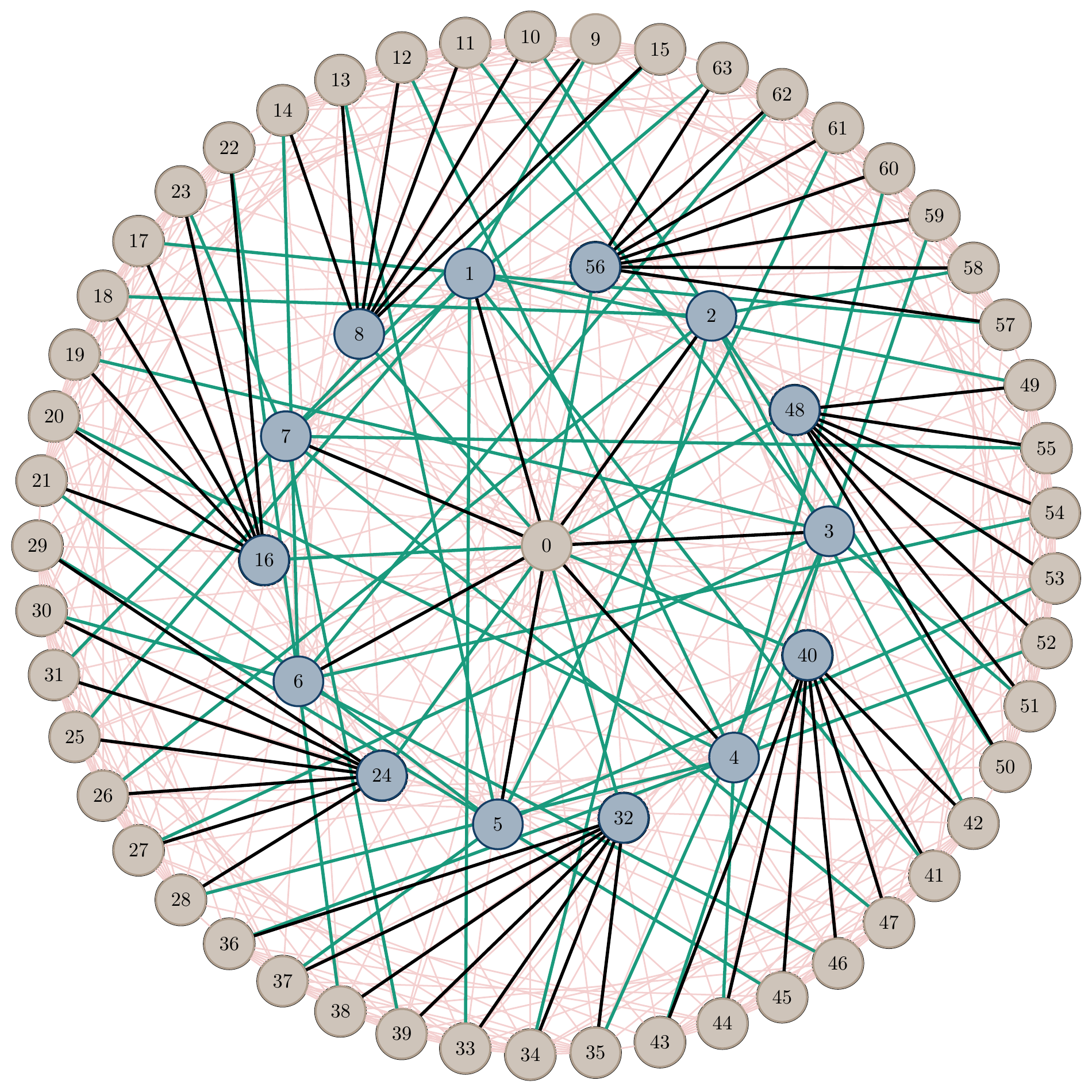}
\caption{\small \label{fig:steane} The structure of encoded Hilbert space for the Steane [[7,1,3]] code and transitions between syndrome subspaces induced by elementary $\sigma_x$ and $\sigma_z$ errors.  The six stabilizer generators for this code are: $S_1 = ZIZIZIZ, S_2 = IZZIIZZ, S_3 = IIIZZZZ, S_4 = XIXIXIX, S_5 = IXXIIXX, S_6 = IIIXXXX$, and the syndrome subspaces (circles) are labeled by the decimal equivalent of their binary syndrome pattern $\bnu$. The central circle with $\bnu=0$ is the codespace. Black lines (green lines) indicate transitions induced by correctable $\sigma_x$ errors ($\sigma_z$ errors). The red lines indicate transitions induced by uncorrectable errors. }
\end{figure}

\begin{figure}
\centering
\includegraphics[scale=0.45]{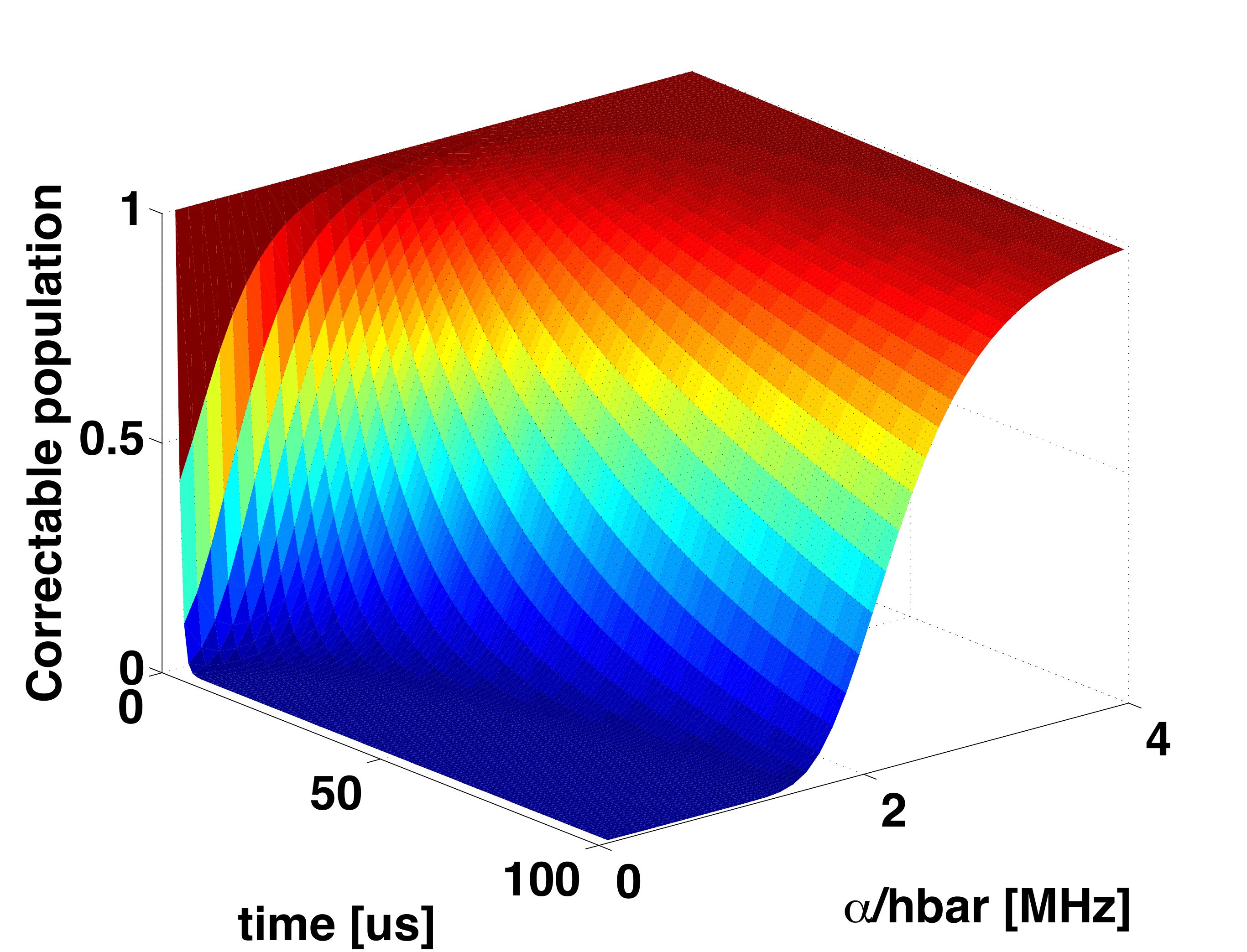}
\caption{\small \label{fig:corr_pop} Population in the correctable subspace as a function of time and $\frac{\alpha}{k_{\rm B} T}$. The average time-independent energy of the logical subspace is taken to be $\bar{\epsilon}(t)=50$kHz $\forall t$. The bath has an Ohmic spectral density with Lorenz-Drude cutoff (see \ref{sec:full_rate_calc} for details), and parameters: $E_R = \hbar(0.1$ MHz), $\gamma=3$MHz and $T = \frac{\hbar}{k_{\rm B}} (1$MHz). The transitions rates in the rate equation \erf{eq:prot_rate_eqn2} are taken in their second Markov approximation and so are time-independent. We confirmed that the population in the correctable subspace has only a weak dependance on the bath parameters that are not varied, as long as the Markovianity condition $E_R/\hbar \ll \gamma$ is met. }
\end{figure}

To illustrate the utility of the rate equation derived above we simulate dynamics under this equation for the example of a single qubit encoded using the Steane $[[7,1,3]]$ code \cite{Steane:1996wn}. The system-bath interaction we consider couples $\sigma_x$ and $\sigma_z$ of each qubit to an environment, and thus induces both bit-flip and phase errors; explicitly:
\beq
H_{\rm SB} = \sum_{i=1}^7 \sigma_x^{(i)} \otimes B_x^{(i)} + \sigma_z^{(i)} \otimes B_z^{(i)}
\label{eq:eg_sb}
\eeq
where $B_{x/z}^{(i)}$ are bath operators. Under this system-bath coupling, the Steane code is capable of correcting all weight one errors and most weight two errors. It is also a perfect code in the sense that every one of the $64$ syndrome pattern identifies a weight one or weight two error.  The syndrome subspaces and connections induced by correctable and uncorrectable errors for this code and system-bath model are shown in Fig. \ref{fig:steane}. We simulate \erf{eq:prot_rate_eqn2} for this code with some typical bath parameters and the results are shown in Fig. \ref{fig:corr_pop}. We see from this graph that when $\alpha/k_{\rm B}T > 2$ the decay of population in the correctable subspace is suppressed heavily. If we extrapolate to long enough times the decay is still exponential, but the decay constant is decreased substantially by the error suppression terms in the Hamiltonian (that are proportional to $\alpha$).

\subsection{Adding local cooling for error correction}
Now we examine the effect of adding local cooling of individual qubits in order to implement a correction mechanism that preferentially populates lower energy states (the code space is the lowest energy state by construction). The cooling is modeled as a strong local coupling of all elementary error operators to a reservoir at low temperature. We will use the term ``reservoir" to refer to the low temperature environment and ``bath" to refer to the uncontrollable environmental degrees of freedom at higher temperature. Thus we add a new interaction and free Hamiltonian to \erf{eq:aqchambar} of the form: $H_{\rm cool} = \sum_{j=1}^{N_e} E_j\otimes F_j + H_{\rm R}$, where $F_j$ are operators in the Hilbert space of the cold reservoir and $H_{\rm R}$ is the free Hamiltonian of the reservoir. This reservoir could physically be a harmonic environment that can be cooled more effectively than other environmental degrees of freedom, or could be ancillary qudits that are actively optically pumped to a low temperature state \cite{Her.You.etal-2009, Young:2012wl}. We will not specify the reservoir details here but simply assume that it is at thermal equilibrium. In this case, one can average over the reservoir degrees of freedom just as we averaged over the bath degrees of freedom in the previous subsections to obtain a rate equation for error path populations in the presence coupling to both environments:
\begin{eqnarray}
\fl \frac{{\rm d}P_{\bnu}(t)}{{\rm d}t} \approx&   \sum_{j~{\rm s.t.} \atop E_jE_{\bnu}~{\rm correctable}}  \left[R^+_{\bnu,j}(t)+S^+_{\bnu,j}(t)\right] P_{\bnu(j)\oplus \bnu}(t)  - \sum_j \left[R^-_{\bnu,j}(t)+S^-_{\bnu,j}(t)\right] P_{\bnu}(t)
\end{eqnarray}
for $\bnu$ correctable, with $R^{\pm}_j(t)$ being the same time-dependent rates as in Eqs. (\ref{eq:r+}) and (\ref{eq:r-}), and the other time-dependent rates (resulting from the coupling to the cold reservoir) being
\begin{eqnarray}
\fl S^+_{\bnu,j}(t) &= \frac{2}{\hbar^2}\Re \left\{ \int_0^t d\tau D_j(\tau) e^{\frac{i}{\hbar}\tau[2\alpha \varpi(j,\bnu)]} \right\} \nonumber \\
\fl S^-_{\bnu,j}(t) &= \left\{
\begin{array}{l l}
\frac{2}{\hbar^2}\Re \left\{ \int_0^t d\tau D_j(\tau) e^{-\frac{i}{\hbar}\tau [2\alpha \varpi(j,\bnu)]} \right\} & \textrm{if $E_j E_\bnu$ is a correctable error} \\
\frac{2}{\hbar^2}\Re \left\{ \int_0^t d\tau D_j(\tau) e^{-\frac{i}{\hbar}\tau [2\alpha \varpi(j,\bnu) + \bar{\epsilon}(t) ]} \right\} & \textrm{if $E_j E_\bnu$ is an uncorrectable error} 
\end{array} \right.
\end{eqnarray}
where $D_{j}(\tau) \equiv \tr_{\rm env}\{\breve{F}_j(\tau)\breve{F}_j(0)\sigma^{{\rm res}}_{\rm eq}\}$ is the quantum correlation function of the reservoir degrees of freedom.

To clarify the effects of cooling, we simplify these rates by taking the second Markov approximation of the bath and reservoir dynamics, and assume $\bar \epsilon(t) \approx \bar \epsilon ~\forall t$. This results in time-independent rates of population transfer and the rate equation
\beq
\frac{{\rm d}P_{\bnu}(t)}{{\rm d}t} \approx   \sum_{j~{\rm s.t.} \atop E_jE_{\bnu}~{\rm correctable}} \gamma(\omega_{\bnu,j}) P_{\bnu(j)\oplus \bnu}(t) - \sum_j \gamma(-\omega_{\bnu,j})  P_{\bnu}(t)
\label{eq:cooled_me}
\eeq
for $\bnu$ correctable, with rates: 
\begin{eqnarray}
\gamma(\omega) =&\frac{2 \mathcal{J}\left(\omega\right)}{\hbar}  \left[\textsf{n}(\omega) + 1\right] + \frac{2 \mathcal{K}\left(\omega\right)}{\hbar}  \left[\textsf{m}(\omega) + 1\right] 
\label{eq:cooled_rates}
\end{eqnarray}
where $\textsf{m}(\omega) = 1/(e^{\beta_R \hbar\omega}-1)$ is the Bose-Einstein distribution at the temperature of the cold reservoir, and $\mathcal{K}(\omega)$ is the spectral density of the reservoir degrees of freedom ($\sf{n}(\omega)$ and $\mathcal{J}(\omega)$ are equivalent quantities for the bath and were defined in section \ref{sec:haqc_eq_0}). For simplicity we assume that this spectral density and distribution of modes is the same for all $j$. Also, 
\beq
\omega_{\bnu,j} \equiv \left\{
\begin{array}{l l}
\frac{2\alpha\varpi(j,\bnu)}{\hbar} & \textrm{if $E_j E_\bnu$ is a correctable error} \\
\frac{2\alpha\varpi(j,\bnu)+ \bar{\epsilon}}{\hbar} & \textrm{if $E_j E_\bnu$ is an uncorrectable error} 
\end{array} \right.
\eeq
As before, this dynamical model tracks population in the correctable subspace and treats any leakage outside this subspace as irrecoverable. Thus it lower bounds the probability of successful computation.

For effective cooling we require two conditions: $\mathcal{K}(\omega) \gg \mathcal{J}(\omega) ~~\forall \omega$ and $\textsf{m}(\omega) \ll \textsf{n}(\omega) ~~ \forall \omega$. The first stipulates that the reservoir is coupled more strongly to the system than the noisy bath (the coupling is still within the weak-coupling regime required for the Born and Markov approximations utilized in deriving the model), and the second stipulates that the reservoir temperature is lower than the bath temperature. Under these conditions, the rates in \erf{eq:cooled_rates} can be approximated by
\begin{eqnarray}
\gamma(\omega) \approx & \frac{2 \left(\mathcal{J}(\omega) + \mathcal{K}(\omega)\right)}{\hbar}
\left(\left[\textsf{m}(\omega) + \frac{\mathcal{J}(\omega)}{\mathcal{K}(\omega)} (\textsf{n}(\omega)-\textsf{m}(\omega))\right] + 1\right) \nonumber
\end{eqnarray}
This describes coupling to an effective reservoir at a slightly higher temperature than the original cooled reservoir. The average thermal occupation of the new effective reservoir is given by the term in the square brackets above:
\beq
\textsf{n}^{\rm eff}(\omega) = \textsf{m}(\omega) + \frac{\mathcal{J}(\omega)}{\mathcal{K}(\omega)} (\textsf{n}(\omega)-\textsf{m}(\omega))
\label{eq:eff_n}
\eeq
and is clearly only perturbatively (in $\mathcal{J}/\mathcal{K}$) higher than the original thermal occupation of the cold reservoir, $\textsf{m}(\omega)$. This average occupation also defines, a possibly energy dependent, effective temperature of the effective reservoir: 
\beq
T^{\rm eff}(\omega) =\frac{\hbar\omega}{k_{\rm B} \ln(\frac{\textsf{n}^{\rm eff}(\omega)+1}{\textsf{n}^{\rm eff}(\omega)})}
\eeq 
Therefore if the temperature of the cooled reservoir can be kept well below the energetic barriers imposed by the EGP, then the population deviation from the codespace can be suppressed even in the presence of the perturbing bath-induced errors, \ie $T^{\rm eff}(\omega_{\boldsymbol{0},j}) \ll \hbar\omega_{\boldsymbol{0},j}/k_{\rm B} ~~\forall j ~\Rightarrow \textsf{n}^{\rm eff}(\omega_{\boldsymbol{0},j}) \approx 0 ~~\forall j~\Rightarrow \gamma(-\omega_{\boldsymbol{0},j}) \approx 0 ~~\forall j$. This is exactly error correction by cooling; \emph{we have constructed an effective reservoir that couples to the appropriate degrees of freedom to quench excitations that represent errors}. 

\section{Conditions for effective error correction and a notion of fault-tolerance in AQC}
\label{sec:ft}
The dynamical equation derived above enables the identification of conditions required for effective error correction by local cooling in AQC. The first condition is of course that we require a protected Hamiltonian that permits restoration of population that has leaked from the codespace (or equivalently, keeps excitations induced by local perturbations localized in space). However, the local nature of the cooling imposes another stringent requirement on successful error correction and long-term error-free operation. This is evident from examining the rate of population arriving and leaving a syndrome space. For the codespace (or a correctable subspace close to the codespace) we want the rate of population leaving as a result of more errors to be smaller than the rate of population returning. This is a minimal condition, because if this were not satisfied then population initialized in the codespace will leak out and become uncorrectable rapidly. That is, we require $\gamma(\omega_{\bnu,j}) > \gamma(-\omega_{\bnu,j})$ if $\mathbf{w}(E_j E_\nu) > \mathbf{w}(E_\nu)$, where $\mathbf{w}(E_\nu)$ is the weight of the error operator $E_\nu$. A consequence of the Kubo-Martin-Schwinger (KMS) condition for a bath at thermal equilibrium \cite{Bre.Pet-2002} is that these transition rates satisfy detailed balance:
\beq
\gamma(-\omega) = e^{-\beta^{\rm eff} \hbar\omega} \gamma(\omega) ~~~ \forall \omega>0 
\eeq
Therefore $\gamma(\omega_{\bnu,j}) > \gamma(-\omega_{\bnu,j})$,  if $\omega_{\bnu, j}=2\alpha\varpi(j,\bnu)/\hbar>0$. But as we see from \erf{eq:varpi}, $\varpi(j,\bnu)$ has no dependence on the weight of the errors and can be positive or negative. The issue is that the energetic cost of an error is dictated by its syndrome pattern (how many stabilizers the error anticommutes with) rather than the Pauli weight of the error. This means that it is possible for \emph{a high-weight error to have a lower energy than a low-weight error}, which is incompatible with error correction by local cooling, which is constructed to drive population towards low energy states. For example, the steady state of \erf{eq:cooled_me} has Boltzmann distributed populations across syndrome subspaces, and this distribution is not favorable unless the lowest energy syndrome subspaces also correspond to the lowest weight error subspaces containing states that are close to the codespace to maximize the probability of a correct decoding. 

However, structuring the energies of syndrome subspaces such that they are ordered by weight of error (or distance from codespace in terms of number of errors) is a challenging problem. 
This can be done be designing the EGP Hamiltonian accordingly, but at the cost of adding an exponential number of energy penalty terms, most of which are high weight; see \ref{sec:thermal_stab}.
In fact, we note that this condition of having a structured energy landscape of error states is exactly the condition required to have a \emph{self-correcting quantum memory} \cite{Dennis:524368,Bac-2006, Bravyi:2009kp, Bombin:2013cv}.
A self-correcting, or resilient, quantum memory is imagined to consist of a lattice finite-dimensional quantum systems whose ground state is degenerate with a finite energy gap to excitations. The ground state degeneracy is stable to weak, local perturbations and these degenerate states are where the quantum information is stored \footnote{We note that a critical part of the definition of a self-correcting quantum memory, the degenerate ground space, is unnecessary in the case of a self correcting AQC implementation.}. The additional key property of self-correcting quantum memories is that they posses \emph{thermal stability}, meaning that if the temperature is below a threshold temperature, $T_c$, excitations from a given ground state created by local perturbations can only result in transitions to the one of the other degenerate ground states in a time exponential in the system size \cite{Alicki:2010wq, Bombin:2013cv, LandonCardinal:2012vp}. This is beneficial since any useful computation or storage is expected to be sub-exponential in system size. 
It is believed that a route to constructing a self-correcting quantum memory is to have a \emph{structured energy landscape} such that each additional error on a state incurs an energy penalty \cite{Bravyi:2009kp}. There are two primary challenges to this approach. The first is that it is currently unknown exactly how the energy penalties should scale with the system size in order to have a thermally stable quantum memory \cite{Michnicki:2012ux}. 
A general condition for thermal stability is complicated by entropic considerations, which reveal that the population of a syndrome subspace is dependent not only on its energetic penalty but also how many error paths lead to the subspace \cite{Bravyi:2009kp}. 
The second challenge is that we only have a few examples of many-body systems where the energy landscape can be structured suitably while at the same time satisfying physical restrictions such as locality, low-weight Hamiltonian terms, and embedding in three or fewer spatial dimensions. 
In fact, an efficient local Hamiltonian construction for such a memory in less than four spatial dimensions is a significant open problem in quantum information \cite{Dennis:524368, Bac-2006, Bravyi:2009kp, Yoshida:2011hj, Bravyi:2011ds}. Therefore we expect that structuring the EGP control Hamiltonian to efficiently implement an energy penalty to erroneous states that depends monotonically on the distance from the codespace (in terms of number of elementary errors) will be a difficult task. 

We emphasize that this is a problem that is orthogonal to the problem of constructing a protected Hamiltonian. The latter, which is constructed by  manipulating $\bar{H}_{\rm AQC}$, is required to prevent coherent delocalization of excitations by the logical Hamiltonian implementing the AQC. The structured energy landscape we refer to in this section, which is constructed by manipulating $H_{\rm C}$, is desirable to prevent environmental/thermal processes from taking the system too far from the codespace through a sequence of errors.

This overlap of the conditions required for a self-correcting quantum memory and a long-term stable AQC stimulates us to ponder on conditions for fault-tolerant AQC. At this stage there is no constructive notion of what it means for an adiabatic quantum computation to proceed in a fault-tolerant manner. However, by drawing an analogy to the case of self-correcting quantum memories it may be possible to posit a limited definition of fault tolerance in AQC.
In the time-continuous, autonomous, models we are considering there is no ``active error correction" implemented by way of gates and measurements, and so the notion of fault-tolerant computing must necessarily be modified from standard circuit model notions stemming from the threshold theorem \cite{mikeandike}. 
In this spirit, a self-correcting quantum memory can be thought of as possessing two phases: a stable phase when $T<T_c$ in which quantum information can be reliably stored in the ground states for a time that scales exponentially with system size, and an unstable phase when $T>T_c$ where thermal fluctuations will corrupt quantum information stored in the degenerate ground space. Such a characterization is analogous to a fault-tolerant quantum computer, which contains a stable operating phase when errors are below threshold and an unstable phase above threshold \cite{Aharonov:2000di} \footnote{We note that although this analogy is informative, there is a critical missing step in it: fault-tolerant quantum computing captures the fact that the computation is resilient to faulty implementations of the gates used for error correction as long as the error rates are below the threshold. As far as we are aware there is no analysis of the stability of self-correcting quantum memories when the non-unitary dynamics that implement cooling and equilibration to the operating temperature, $T<T_c$, contain small errors.}. 
This suggests that we define an \emph{environmental-fault-tolerant AQC implementation} as one capable of executing AQC evolution and error correction at once, while also possessing thermal stability in the sense that crossing the boundary from correctable to uncorrectable states through environmentally induced processes takes a time that scales exponentially with encoded system size. Note that this definition of fault-tolerance is limited since it does not capture the system's susceptibility to failure modes that are not induced by the system-bath coupling such as diabatic errors and failure to implement the correct final Hamiltonian. The dynamical model developed in this paper can be used to formulate sufficient conditions for such fault-tolerant AQC. The analysis in section \ref{sec:error_corr} demonstrates how to implement error correction by local cooling. The remaining step, of stipulating conditions on $H_{\rm C}$ that produce a favorable energy landscape is treated in \ref{sec:thermal_stab}, where we use a simplified dynamical model to numerically explore the effect of various energy landscapes on thermal stability.

Finally, we note that the monotonic energy landscape requirement comes from the fact that our method for entropy reduction is the restrictive local cooling model. That is, a local cooling operation can only act on information collected from a local neighborhood of the whole system (one qubit in the case above), whereas the optimal decoding and correction operation must correlate and compare all the stabilizer measurements across the whole lattice of qubits. Therefore, it may be possible to replace the structured landscape requirement with a continuous-time physical implementation of a near-optimal decoder that acts upon a large fraction of the stabilizer syndrome values. It is unclear what such a physical implementation would look like but recent constructions of physical realizations (as opposed to realizations on a digital processor) of decoders for classical codes may suggest a route forward \cite{Pavlichin:2013wh}.

\section{Discussion}
\label{sec:disc}

We have analyzed the dynamics of stabilizer code encoded adiabatic quantum computing in the presence of a weakly coupled perturbing environment. We constructed an open system model describing error dynamics for encoded AQC that enabled the unification of previously proposed techniques for error suppression, energy gap protection and dynamical decoupling, under the same dynamical framework. The model elucidates the mechanisms behind error suppression for both techniques and allows calculation of rates of leakage from the codespace. We note that our dynamical model is applicable to any situation where stabilizer encoding is used to protect evolving ground states. Therefore, it could also be useful for modeling the effects of noise on encoded quantum simulation.

Then we extended the model to encompass error correction in encoded AQC by local cooling. The steps taken in deriving this dynamical model clarify the essential physical properties of the system and environment required for the validity of popular Markovian rate models of errors and perturbations used in quantum computing. In particular, we identified several requirements for error correction to be successful in the adiabatic model of quantum computation:
\begin{enumerate}
\item The stabilizer code structure should be imposed by energy gap protection as opposed to dynamical decoupling. The latter does not impose real energy penalties on error states and therefore is not compatible with cooling based error correction. 
\item A choice of logical operators in the encoded adiabatic Hamiltonian, $\bar{H}_{\rm AQC}$, is needed that ensures that local excitations remain localized in space and energy. That is, the eigenstates of the logical Hamiltonian in the codespace should remain eigenstates after the occurrence of an error that promotes the state into a error syndrome subspace. These Hamiltonians, termed \textit{protected Hamiltonians}, can be constructed using the freedom in defining the logical operators of a stabilizer code.
\item Error correction can be implemented by coupling local systems (qubits) to a thermalizing cold reservoir with temperature less than the EGP penalty. This reservoir could be implemented by selectively coupling multi-level ancilla systems and optical pumping \cite{Young:2012wl}.
\end{enumerate}
Finally, we also considered the conditions necessary for long-term, stable (``fault-tolerant") operation of an AQC with error correction implemented through local cooling.

Unfortunately these requirements for error correction and long-term stable operation are extremely demanding experimentally. All the requirements identified above, except for the ability to couple to a cold reservoir, can only be implemented by increasing the weight of the Hamiltonian of the encoded system. Therefore although we have identified the requirements for performing error correction within an adiabatic model of quantum computation, these requirements are likely too stringent to be practical, at least in the near-term.

\section*{Acknowledgements} 
\label{sec:acknowledgements}
We acknowledge useful discussions with Sandia's AQUARIUS Architecture team, especially with Robin Blume-Kohout, Anand Ganti, and Andrew Landahl. M.S. also acknowledges useful discussions with David Poulin on the topic of self-correcting quantum memories. This work was supported by the Laboratory Directed Research and Development program at Sandia National Laboratories. Sandia is a multi-program laboratory managed and operated  by Sandia Corporation, a wholly owned subsidiary of Lockheed Martin Corporation, for the United States Department of Energy's National Nuclear Security Administration under contract DE-AC04-94AL85000.


\section*{References}
\bibliographystyle{unsrt}
\bibliography{aqc_dynamics}

\newpage

\appendix
\section[\hspace{2cm} Full rate calculation for Ohmic spectral density]{Full rate calculation for Ohmic spectral density}
\label{sec:full_rate_calc}
The example environment considered in the main text is a classical bath with exponentially decaying correlation. Here we generalize this to a true quantum environment and explicitly demonstrate that the controlled suppression of population leakage from the codespace holds in this case too. Consider a damped harmonic environment with an Ohmic spectral density with Lorentz-Drude regularization: 
\beq
\mathcal{J}(\omega) =  2 E_R  \frac{\gamma \omega}{\omega^2 + \gamma^2},
\label{eq:lorentz-drude}
\eeq
where $E_R$ is the \textit{reorganization energy}, which quantifies the total system-environment coupling strength, and $\gamma$ is the inverse of the environment correlation timescale. The quantum correlation function for an environment with such a spectral density is:
\begin{equation}
C(t) = i2\hbar E_R \gamma \left( \frac{1}{e^{i\beta\hbar\gamma}-1}\right)e^{-\gamma t} - \sum_{\kappa=1}^\infty \frac{4 E_R \gamma}{\beta}\frac{\nu_\kappa}{\gamma^2-\nu_\kappa^2}e^{-\nu_\kappa t} \nonumber
\end{equation}
where $\beta=1/k_{\rm B}T$ is the inverse temperature and $\nu_\kappa \equiv \frac{2\pi \kappa}{\beta\hbar}$ are the \textit{Matsubara} frequencies. For moderate to high temperatures the terms in the summand decay quickly and it is customary to truncate the sum at finite $\kappa$. Assuming that the error suppression technique is EGP and computing the rates in the population master/rate equation (\erf{eq:pop_eqn}) yields:
\begin{eqnarray}
\fl r^\pm_j(t) = \frac{ 2E_R \gamma \csc(\frac{\beta \hbar \gamma}{2})}{\hbar}\frac{[ a_0^\pm -\gamma e^{-\gamma t}\cos( \pm \omega_j t - \frac{\beta \hbar \gamma}{2}) \pm \omega_j e^{-\gamma t}\sin(\pm \omega_j t - \frac{\beta \hbar \gamma}{2}) ] }{\omega_j^2 + \gamma^2} \nonumber \\
- \sum_{\kappa=1}^\infty \frac{8E_R \gamma}{\beta\hbar^2} \frac{\nu_\kappa}{\gamma^2 - \nu_\kappa^2} \frac{[ \nu_\kappa - \nu_\kappa e^{-\nu_\kappa t}\cos(\omega_j t ) + \omega_j e^{-\nu_\kappa t}\sin(\omega_j t) ] } {\omega_j^2 + \nu_\kappa^2}
\end{eqnarray}
where $\omega_j=2\alpha w_j/\hbar$, $a_0^\pm= \left[\gamma \cos(\frac{\beta \hbar \gamma}{2}) \pm \omega_j \sin(\frac{\beta\hbar\gamma}{2})\right]$. As in the case of a classical model of the environment, the suppression of these transition rates is achieved by two mechanisms: (i) the suppression term $2\alpha w_j$ in the denominator decreases the overall rate of population leakage, (ii) the same term increases the oscillation frequency of the sinusoidal functions in the numerator, thus decreasing the magnitude of integrals of $r^\pm_j(t)$. 

\begin{figure}
\centering
\includegraphics[scale=0.5]{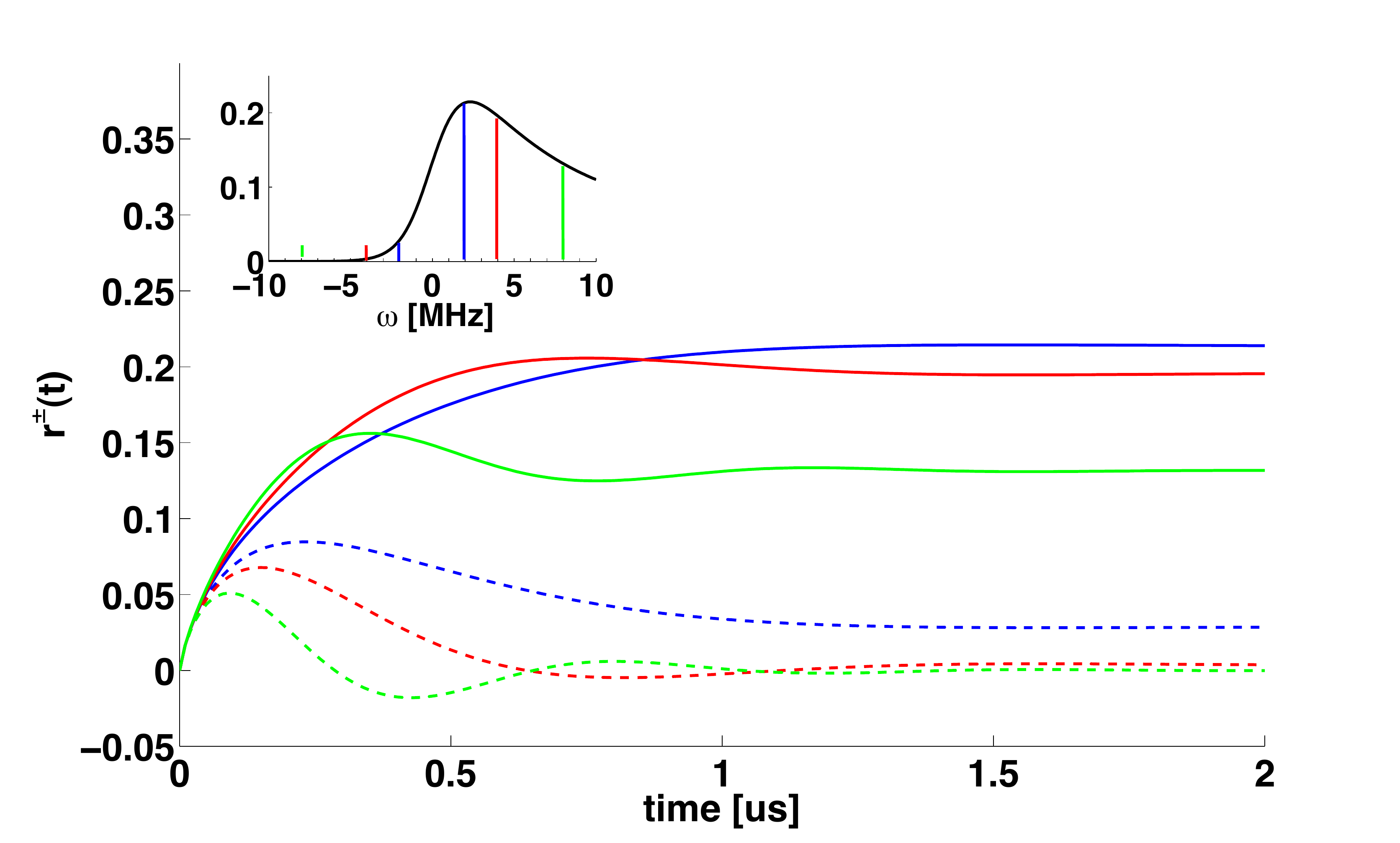}
\caption{\small \label{fig:rates} Time dependent transition rates between codespace and error space for a bath with Ohmic spectral density. Solid lines are $r_j^+(t)$ and dashed lines are $r_j^-(t)$. The error weight is assumed to be $w_j=1$, and hence $\omega_j= 2\alpha/\hbar$. The bath parameters used are $\gamma=3$MHz, $E_R=\hbar (0.1$ MHz), $T=(\hbar/k_{\rm B})(1 $MHz). Three values of energy penalty are shown: $\alpha/\hbar=1$MHz (blue), $\alpha/\hbar=2$MHz (red), $\alpha/\hbar=4$MHz (green). The inset shows $\frac{2}{\hbar^2} \Re {\sf C}_j(\omega)$, where ${\sf C}$ is the one-sided Fourier transform of $C(t)$. In a second-Markov approximation, this is the quantity that determines the rates and we have indicated with colored lines (the color-to-$\alpha$ mapping is the same as in the main figure) the points at which this quantity is sampled.  At long times, $r_j^{\pm}(t)$ approaches $\frac{2}{\hbar^2} \Re {\sf C}_j(\omega_j)$. This intuitively explains the suppression of $r^\pm$; at this temperature the Ohmic spectrum dictates that $\Re {\sf C}_j(\omega)$ quickly decays when $\omega \gtrsim 2$MHz and is negligible when $\omega\lesssim-5$MHz. Note that when $\omega_j>\gamma$, $r^-(t)$ can become $<0$ at short times. This is a well-known problem with time-dependent Redfield rates: the transient behavior can lead to negative rates as a result of the theory being invalid for very short times. There are various solutions to this problem, \eg \cite{Suarez:1992vv}, but we will not concern ourselves with these here since the long-time behavior is of most interest to us and these transient effects are negligible for long-time behavior.}
\end{figure}

We plot these rates for some sample parameters in Figure \ref{fig:rates}.

\section[\hspace{2cm} Simplified thermal stability analysis]{Simplified thermal stability analysis}
\label{sec:thermal_stab}
In the main text we reduced the open system dynamics of an encoded AQC with a protected logical Hamiltonian to a Markov random walk description with time-independent rates when all Markov approximations of the environment are made. Such a description is also valid for lattice based quantum memories encoded using a stabilizer code Hamiltonian (\eg the abelian toric code \cite{Kitaev:2003ul}). This description can be used to analyze properties required of the quantum code (which translates to properties required of the energy penalty enforcing Hamiltonian, $\hc^{\rm EGP}$) for the type of thermal stability required for long-term operation of the quantum memory or AQC. We define a thermally stable system in this context as one in which the time taken for a sequence of physical errors (stemming from a system-bath coupling) to build to an uncorrectable logical error scales exponentially in the ``system size". For quantum memories this system size is often defined as the number of physical qubits (\eg the number of qubits in a 2D lattice encoding $k$ qubits in its ground state). Alternatively, if one is building a system for computation from $n_l$ single qubits, each encoded in a concatenated code, then it is desirable for the system to remain correctable for a time exponential in the number of logical qubits, $n_l$, since any efficeint computation will execute for a time less than this. Hence, in such systems thermal stability is the exponential scaling of population in the correctable subspace with the number of logical qubits. The question we address in this Appendix is what kind of scaling of the energy penalty Hamiltonian with $n_l$ is required for this type of thermal stability?

As discussed in the main text, this type of thermal stability is intimately related to the type of energy landscape the errors/excitations experience. The most straightforward way to do such a thermal stability analysis is to simulate error dynamics, using for example a Markov chain Monte Carlo algorithm, with physically motivated rates calculated from the derived expressions. In this appendix, we take an alternate route and use a key simplification to relate this problem to a one-dimensional (1D) Markov chain hitting time problem whose analytical solution can be derived easily. 

Consider a non-degenerate code whose syndrome subspace populations dynamics are described by \erf{eq:prot_rate_eqn2}, and furthermore, consider the full Markov limit where the rates of population transfer become time-independent and only dependent on the energy difference between the syndrome subspaces. To map the dynamics to a discrete-time Markov random walk we choose a time discretization $\Delta t$ and rewrite the dynamics as:
\begin{eqnarray}
P_{\bnu}[n] \approx   \sum_{j~{\rm s.t.} \atop E_jE_{\bnu}~{\rm correctable}} \pi_{\bnu,j} P_{\bnu(j)\oplus \bnu}[n-1] + \left( 1 - \sum_{j} \chi_{\bnu,j} \right) P_{\bnu}[n-1]  
\label{eq:disctime_prot_rate_eqn}
\end{eqnarray}
for $\bnu$ correctable. $P_{\bnu}[n] \equiv P_{\bnu}(n\Delta t)$, and the coefficients are probabilities defined as:
\bqa
\pi_{\bnu,j} &=& \Delta t \frac{2 \mathcal{J}(\omega_{\bnu,j})}{\hbar}[\textsf{n}(\omega_{\bnu,j})+1] \nn \\
\chi_{\bnu,j} &=& \Delta t \frac{2 \mathcal{J}(-\omega_{\bnu,j})}{\hbar}[\textsf{n}(-\omega_{\bnu,j})+1]
\eqa
Recall that $\omega_{\bnu,j}$ is the frequency difference between the syndrome subspaces connected by the particular transition. The spectral density and temperature in the above expression can be properties of the environment, or renormalized versions that result from coupling to a cold reservoir (\eg \erf{eq:eff_n}). This discretization of the dynamics makes the following analysis easier and we shall see below that the particular choice of $\Delta t$ does not affect the results.

\begin{figure}
\centering
\includegraphics[scale=0.62]{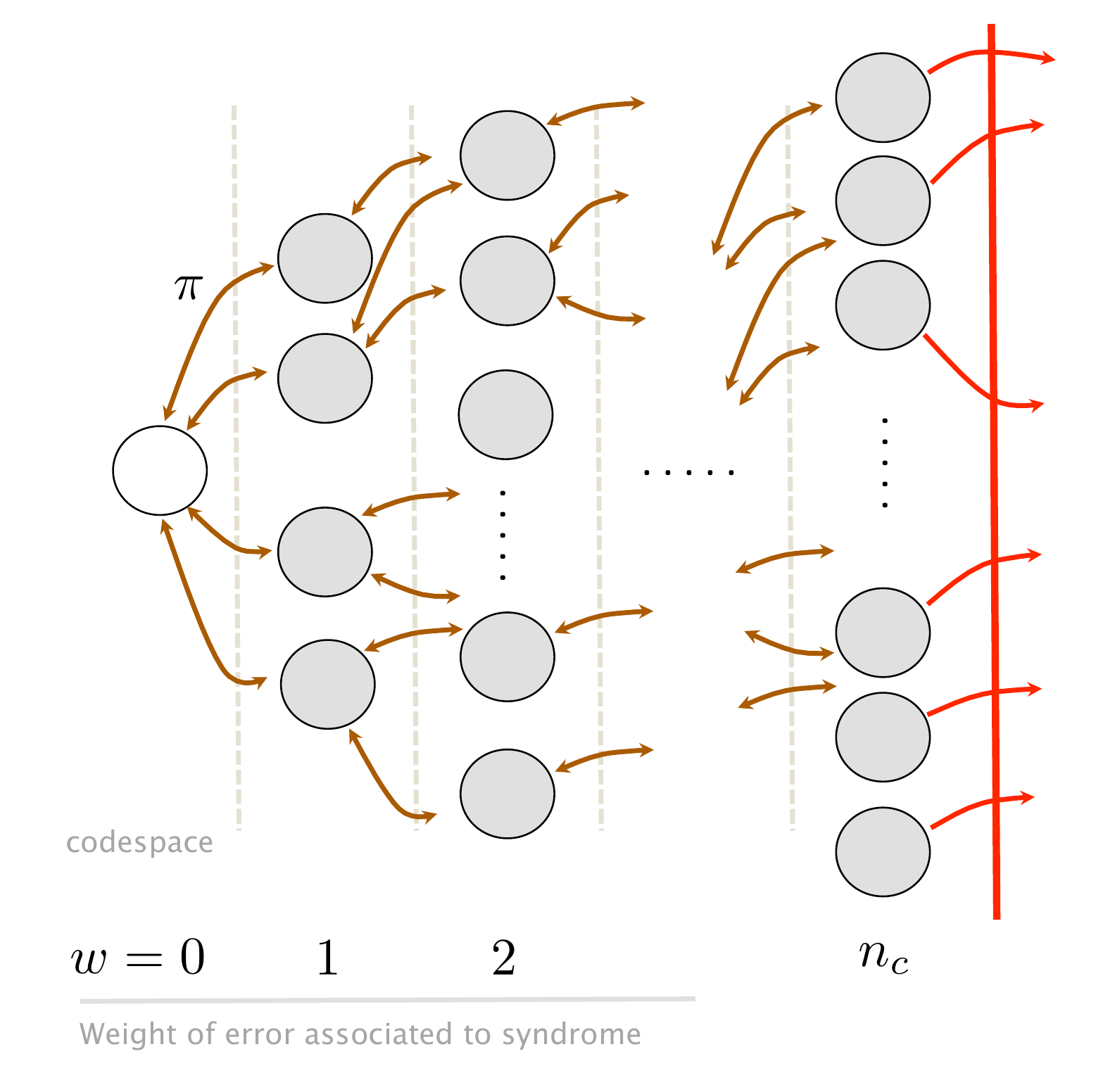}
\caption{\small \label{fig:app_nondegen} A redrawing of the structure of encoded Hilbert space for an arbitrary non-degenerate code with distance $d$ and number of correctable errors $n_c = \lfloor \frac{d-1}{2} \rfloor$. The syndrome subspaces are organized into columns with column $w$ containing all syndrome subspaces with errors of weight $w$ associated to their states. Arrows indicate transitions between states, and transitions across the red line create logical errors.}
\end{figure}

In Fig. \ref{fig:app_nondegen} we draw a time-discretized version of the encoded state space diagrams in the main text (\eg Fig. \ref{fig:steane}). Here, all states with a certain error weight are grouped together in columns. There are $\scriptsize \left(\begin{array}{c} N_e \\ w \end{array}\right)$ states in column $w$, and all states in column $w$ have $w$ transitions back to column $w-1$ and $(N_e-w)$ transitions forward to column $w+1$. Crossing the red line with any transition results in an uncorrectable state. Since we are focusing on non-degenerate codes the red transitions actually lead back to the drawn syndrome subspaces. However, in the following we will treat any transition across the correctable-uncorrectable as leakage and not track this population. 

At this point we utilize a critical simplification and equate all transition probabilities from column $w$ to $w+1$ and denote this probability by $\pi_{w\rightarrow (w+1)}$. Similarly, all transition probabilities from column $w$ to $w-1$ are considered to be the same and denoted by $\pi_{w\rightarrow (w-1)}$. This simplification can result from a choice of EGP Hamiltonian that enforces an energy penalty on states that depends only the weight of the error assigned to the syndrome value of that state; \ie a Hamiltonian of the form:
\beq
\hc^{\rm EGP} = \alpha \sum_{w=1}^{\lfloor \frac{d-1}{2} \rfloor} \sum_{\mathbf{w}(E_\nu)=w} \delta(w) E_\bnu \mathbf{P} E_\bnu
\label{eq:egp_landscape}
\eeq
where $\alpha>0$ and $\delta(w)$ is a scaling factor for states with weight $w$ errors. We also define $\Delta_w \equiv \delta(w)-\delta(w-1)$. It is clear that this Hamiltonian is extremely high weight and therefore not practical. However, the required simplification can also result from a replacement of all transition probabilities from column $w$ to $w+1$ by their maximum; \ie $\pi_{w\rightarrow (w+1)} = \max \{\pi_{\bnu,j} | \mathbf{w}(E_\nu)=w ~{\rm and}~ \mathbf{w}(E_j E_\nu)=w+1\}$. Then $\pi_{(w+1)\rightarrow w}$ is the probability of the corresponding backward transition and $\Delta_w$ is the energy difference between the states that are connected by this transition. In this case the following can be viewed as a worst-case analysis where $\pi_{w\rightarrow (w+1)}$ is arrived at by choosing the maximum transition probability.

\begin{figure}
\centering
\includegraphics[scale=0.7]{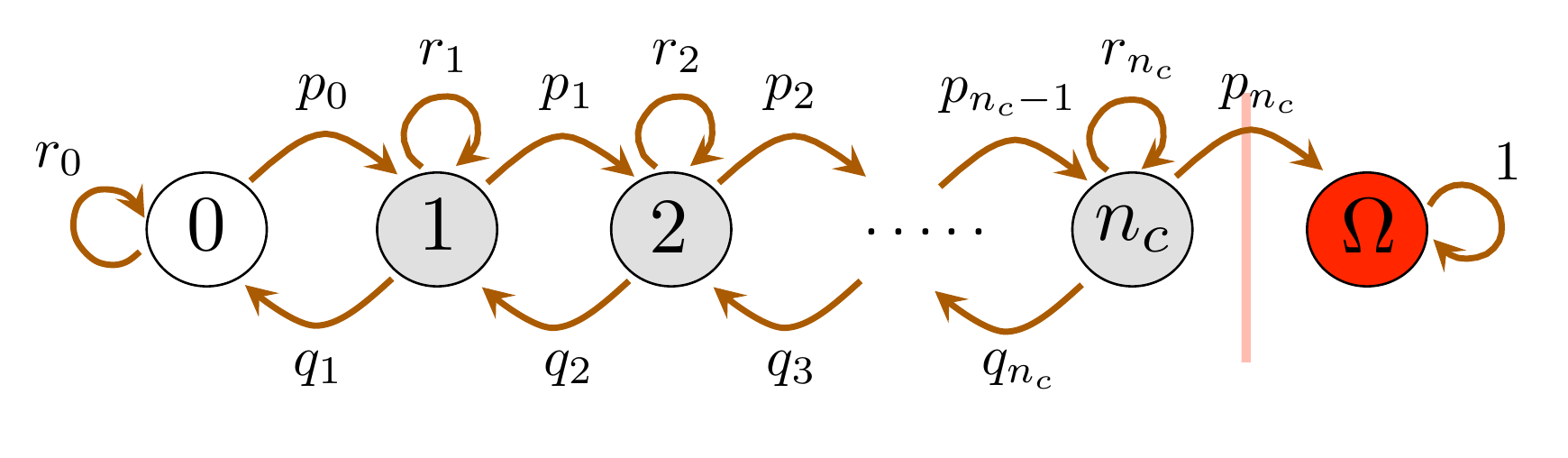}
\caption{\small \label{fig:app_mc} A 1D Markov chain representation of \ref{fig:app_nondegen} achievable when states lying in each column can be lumped. $p_w$ and $q_w$ are transition probabilities and $r_w = 1-p_w-q_w$. Note that $q_0=q_{\Omega}=0$.}
\end{figure}

Under the above simplification the Markov chain in Fig. \ref{fig:app_nondegen} satisfies conditions for \textit{strong lumpability} \cite{Filliger:2008ts} and we can group all states in a given column into one and recover an equivalent one-dimensional (1D) Markov chain, as in Fig. \ref{fig:app_mc}. Any transition across the correctable-uncorrectable boundary is treated as a transition into an absorbing state, $\Omega$. The forward and backward transition rates in this 1D Markov chain are explicitly:
\bqa
p_w &=& {\scriptsize \left(\begin{array}{c} N_e \\ w \end{array}\right)} (N_e-w) \pi_{w\rightarrow (w+1)} \nn \\
q_w &=& {\scriptsize \left(\begin{array}{c} N_e \\ w \end{array}\right)} w ~\pi_{w\rightarrow (w-1)} 
\eqa
with $\pi_{w\rightarrow (w+1)} = \Delta t \frac{2\mathcal{J}(-\alpha \Delta_{w+1})}{\hbar}[\mathsf{n}(-\alpha \Delta_{w+1})+1]$ and $\pi_{w\rightarrow (w-1)} = \Delta t \frac{2\mathcal{J}(\alpha \Delta_w)}{\hbar}[\mathsf{n}(\alpha \Delta_w)+1]$. From the detailed balance condition $\frac{\pi_{w\rightarrow (w-1)}}{\pi_{(w-1)\rightarrow w}} = e^{\beta \alpha\Delta_w}$, we know that $\frac{q_w}{p_{w-1}} = e^{\beta \alpha\Delta_w}$.

Fig. \ref{fig:app_mc} describes a 1D Markov chain with one reflecting and one absorbing boundary condition. We are ultimately interested in the mean hitting time to the absorbing state (or equivalently the mean absorption time) of a random walk that is initialized at the $w=0$ state. This can be easily calculated using a first step analysis \cite{vanMieghem:2009wa}, and results in:
\beq
\eta_0 \equiv \mathbb{E}\{t_{\rm abs} | P_{\bnu=0}[0]=1\} = \sum_{k=1}^{n_c} \frac{\Delta t}{p_{k-1}}\left(1 + \sum_{n=1}^{n_c-k} \prod_{m=1}^n \frac{q_{m+k-1}}{p_{m+k-1}}\right)
\eeq
where $t_{\rm abs}$ is the stochastic absorption time. Note that there is no real $\Delta t$ dependence in this expression since $p_w \propto \Delta t$, and hence the result is not dependent on the choice of discretization time. Rewriting this expression in terms of the ratios $\frac{q_w}{p_{w-1}}$, we get
\beq
\eta_0 = \sum_{k=1}^{n_c} \left[ \frac{1}{\tilde{p}_{k-1}} + \sum_{n=1}^{n_c-k} \frac{1}{\tilde{p}_{n+k-1}}\left( e^{\beta \alpha \sum_{m=1}^n \Delta_{m+k-1}}\right) \right]
\label{eq:hitting}
\eeq
where $\tilde p_w = p_w/\Delta t$. This quantity seems to grow exponentially with each positive energy barrier ($\Delta_w>0$) as expected, but also contains combinatorial factors (in $\tilde p_w$) that encode the entropic contribution to the average hitting time. 

\begin{figure}
\centering
\includegraphics[scale=0.63]{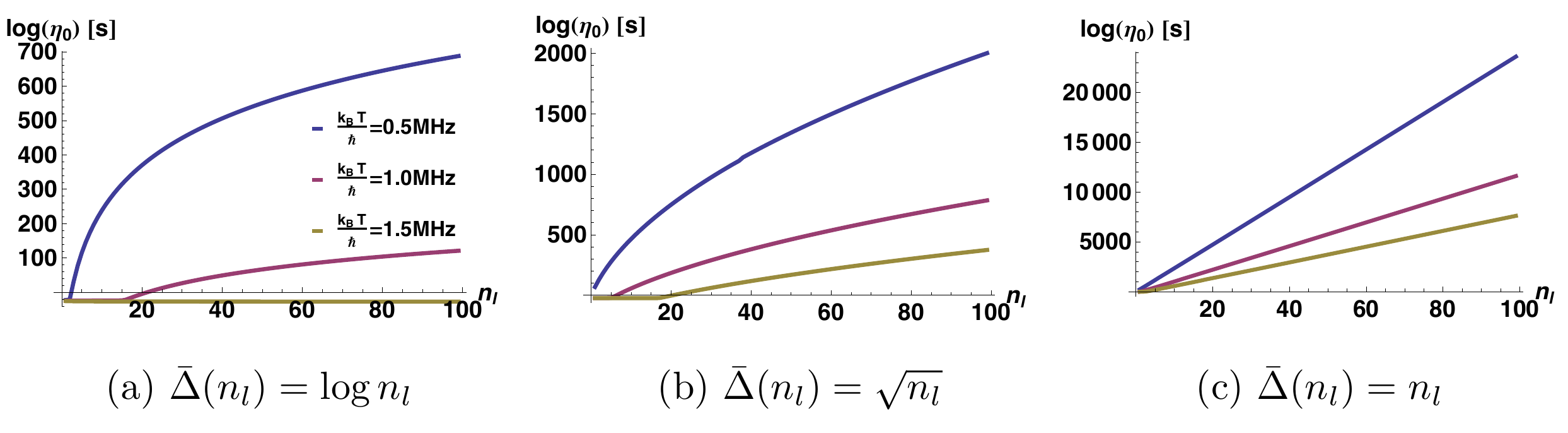}
\caption{\small \label{fig:hitting_n} Log of the mean hitting time versus number of logical qubits ($n_l$) for three types of scaling for the energetic barrier $\Delta_w$: (a) logarithmic scaling, (b) square root scaling, and (c) linear scaling with $n_l$. The penalty energy scale is set by $\alpha/\hbar=3$MHz. The encoding for each logical qubit is the Steane code to 4 levels of concatenation, and the three curves in each graph indicate behavior for three different bath temperatures. The bath was chosen to have Ohmic spectral density with Lorentz-Drude regularization (\erf{eq:lorentz-drude}), with parameters: $E_R=\hbar (0.1$ MHz), $\gamma=200$MHz (the regularization rate $\gamma$ is chosen to be very large to be consistent with the second Markov approximation of the rates). }
\end{figure}

Fig. \ref{fig:hitting_n} shows the results of numerical solutions of \erf{eq:hitting} for a system encoded in a concatenated $[[7,1,3]]$ Steane code with elementary errors $\sigma_x$ and $\sigma_z$ on each qubit -- \ie the system-environment interaction is as \erf{eq:eg_sb} in the main text (see figure caption for bath parameters). Recall that at a concatenation level of $t$, the number of physical qubits per logical qubit is $7^t$ and the distance of the code is $3^t$. Here we consider $t=4$ and assume all errors up to weight $n_c = \lfloor \frac{3^4-1}{2} \rfloor$ can be corrected. Fig. \ref{fig:hitting_n} shows how the average hitting time scales with different energy landscapes imposed by choice of the energy barriers. All energy barrier steps are taken to be the same for simplicity -- \ie $\Delta_w = \bar{\Delta} ~ \forall w$ -- and we consider how $\bar{\Delta} \rightarrow \bar{\Delta}(n_l)$ has to scale with the number of logical qubits $n_l$ (each logical qubit is encoded in the concatenated Steane code) to achieve various scalings of average hitting time with $n_l$. In (a) the cost of each additional error scales logarithmically in $n_l$, while in (b) and (c) this cost scales as a small power and linearly, respectively. 

\begin{figure}
\centering
\includegraphics[scale=0.35]{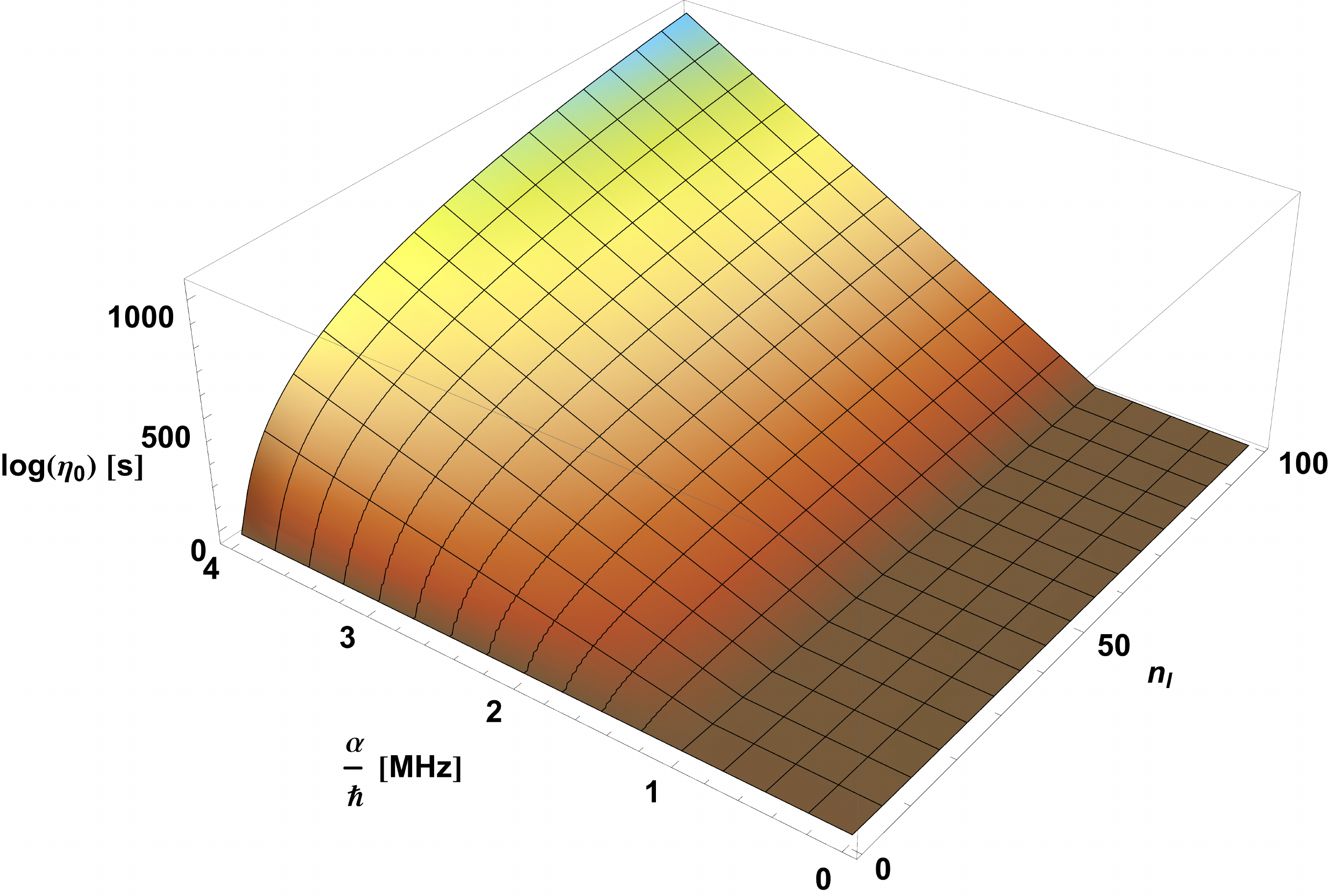}
\caption{\small \label{fig:hitting_b} Log of the mean hitting time as a function of $\alpha$ and $n_l$ (for $\bar{\Delta} = \log(n_l)$). The bath temperature is $k_B T/\hbar = 0.5$MHz and all other parameters are as in Fig. \ref{fig:hitting_n}. The similarity of the scaling of log hitting time and $\bar{\Delta}$ (both with $n_l$) is only present after $\alpha \gtrsim k_B T$.}
\end{figure}

Two observations to make from \ref{fig:hitting_n} are: 
\begin{enumerate}
\item As the number of (logical and physical) qubits is increased, log of the average hitting time scales in the same manner as the scaling of $\bar{\Delta}(n_l)$. Hence to achieve exponential scaling of average hitting time -- \ie exponential scaling of the failure time for a computation or memory -- we require the energy cost of each additional error to scale linearly in system size for this concatenated code approach.
\item This scaling of $\log(\eta_0)$ as $\bar{\Delta}(n_l)$ is only recovered once the energy barrier is above a certain level, as evidenced by the initial flat trend of $\log(\eta_0)$ when $k_B T/\hbar = 1.0$MHz and $1.5$MHz in Fig. \ref{fig:hitting_n}(a) and (b). In fact, this behavior is more convincingly seen when we examine the average hitting time as a function of the number of logical qubits and the energy barrier scale $\alpha$, as in Fig. \ref{fig:hitting_b} (plotted only for $\bar{\Delta}(n_l)=\log(n_l)$). This plot shows that the scaling of $\log(\eta_0)$ with $n_l$ turns over from a flat line to one that behaves as $\bar{\Delta}(n_l)$ at some critical value of $\alpha$ (which is about $1$MHz) in this case. That is, when $\alpha$ is below this critical value the hitting time remains constant, even if $\Delta_l$ scales with $n_l$.
\end{enumerate}

To understand these two interesting aspects of the average hitting time we further analyze the analytical expression for $\eta_0$. We begin with the form of \erf{eq:hitting} when $\Delta_w = \bar{\Delta}$ is independent of $w$: 
\beq
\eta_0 = \sum_{k=1}^{n_c} \sum_{n=0}^{n_c-k} \frac{e^{\lambda n}}{\tilde{p}_{n+k-1}} = \frac{1}{\pi_{\bar{\Delta}}} \sum_{k=1}^{n_c} \sum_{n=0}^{n_c-k} \frac{e^{\lambda n}}{\scriptsize \left(\begin{array}{c} N_e \\ n+k \end{array}\right)(n+k)}
\label{eq:hitting_constant_D}
\eeq
where $\lambda \equiv \beta \alpha \bar{\Delta}$, and we have used the fact that in this case of independent energy barrier steps, $\tilde{p}_w = {\scriptsize \left(\begin{array}{c} N_e \\ w \end{array}\right)(N_e-w)}\pi_{\bar{\Delta}}$, with $\pi_{\bar{\Delta}} = \left[ j \left( \frac{e^{-\lambda}}{1-e^{-\lambda}}\right)\right]$, with $j_{\bar \Delta} \equiv \frac{2 \mathcal{J}(\alpha\bar{\Delta}/\hbar)}{\hbar}$. Now, we will lower bound this quantity by replacing $n+k$ in the summand by its maximum value, $n_c$, and use 
\beq
\frac{1}{\scriptsize \left(\begin{array}{c} N_e \\ n+k \end{array}\right)(n+k)} \geq \frac{1}{\scriptsize \left(\begin{array}{c} N_e \\ n_c \end{array}\right)n_c}
\eeq
when $N_e \gg n_c$. To see the validity of this bound note that for a concatenated $[[n,k,d]]$ code $N_e \propto n^t$ and $n_c = \lfloor \frac{d^t-1}{2} \rfloor$, and since $n>d$ by the quantum Singleton bound \cite{mikeandike} $N_e \gg n_c$ for $t\geq 2$. For instance, the $[[7,1,3]]$ Steane code concatenated to $t=4$ levels, yields $N_e \propto 2401$ and $n_c = 20$. Using this inequality and the expression for $\pi_{\bar{\Delta}}$ above, we get the lower bound
\bqa
\eta_0 \geq \eta_{\rm bound} \equiv  \frac{1}{\scriptsize \left(\begin{array}{c} N_e \\ n_c \end{array}\right) j_{\bar \Delta}} ~ \left[ \frac{e^{\lambda(n_c+1)}+1}{n_c(1-e^{-\lambda})} - 1\right]
\label{eq:eta_bound}
\eqa

Consider this bound in the regime valid for the simulations in Fig. \ref{fig:hitting_n}, $\lambda = \frac{\alpha \bar{\Delta}}{k_B T} \gg 1$, which is when each energy barrier step is larger than the thermal energy. The log of \erf{eq:eta_bound} in this regime is well approximated by
\beq
\log(\eta_{\rm bound}) \approx \lambda (n_c+1) - N_e H(\frac{n_c}{N_e}) - \log(j_{\bar \Delta}) - \log(n_c)
\eeq
where $H(p)\equiv -p\log p - (1-p)\log(1-p)$ is the binary entropy function (the $\log$ is base 2) and we have used the approximation $\log {\scriptsize \left(\begin{array}{c} n \\ k \end{array}\right)} \approx n H(k/n)$, valid when $n\gg 1$ and $k\gg 1$. For the concatenated code example considered in Fig. \ref{fig:hitting_n}, $\lambda = \beta \alpha \bar{\Delta}(n_l)$ and $N_e = 2 (7^4) n_l$, and substituting these values,
\beq
\fl \log(\eta_{\rm bound}) \approx \beta \alpha \bar{\Delta}(n_l) (n_c+1) - 2 (7^4) n_l H(\frac{n_c}{2 (7^4) n_l}) - \log( \frac{2 \mathcal{J}(\frac{\alpha\bar{\Delta} (n_l)}{\hbar})}{\hbar} ) - \log(n_c)
\label{eq:log_n_bound_approx}
\eeq
This expression explains both of the observations listed above if we examine its behavior with respect to $n_l$. We see that for small $n_l$ the second entropic factor dominates unless $\bar{\Delta}(n_l)$ is linear in $n_l$. This is the reason for the small, almost constant hitting time for small $n_l$ on some curves in \ref{fig:hitting_n}(a) and (b). However, as $n_l$ grows the second entropic factor becomes logarithmic in $n_l$ since $xH(1/x) \rightarrow \log(x)$ as $x\rightarrow \infty$, and hence the first factor will dominate as long as $\bar{\Delta}(n_l)$ grows at least logarithmically in $n_l$. Precisely where this turnover happens is dictated by $\beta, \alpha$ and parameters of the code. This shows that the scaling of the energy barrier step with logical qubits eventually dictates the hitting time behavior, $\eta_0(n_l) \sim e^{\bar{\Delta}(n_l)}$, as long as this scaling is logarithmic or higher, and $\lambda \gg 1$. This suggests that thermal stability, defined as exponential scaling of hitting time with $n_l$, is achievable only when the energy barrier steps $\bar{\Delta}$ grow linearly in $n_l$, a demanding requirement.

One can follow a similar line of analysis to show the following scaling properties of $\eta_0$: \textit{(i)} $\eta_0$ is doubly exponential in the concatenation level $t$, and \textit{(ii)} in the $\lambda \ll 1$ regime, the hitting time decays with $n_l$ even if $\bar{\Delta}(n_l)$ scales linearly in $n_l$. 

In section \ref{sec:eg_classical_noise} we pointed out that the rate of leakage from the codespace can be exponentially suppressed if the spectral density of the bath decays exponentially and the EGP energy penalty can be scaled linearly with $n_l$ (see also \cite{Jordan:2006jb}). We seem to have arrived at the same requirement on energy penalties in this analysis of thermal stability. However, the key difference here is that the exponential scaling is achieved with no strong assumption on the bath spectral density. The dependence of \erf{eq:log_n_bound_approx} on the spectral density at $\bar \Delta$ is logarithmic and therefore it will not dominate the behavior unless the spectral density \textit{increases} exponentially with energy, which is an unphysical dependence. Therefore, the addition of the protected Hamiltonian and concatenated error correction mechanisms allow the exponential (in $n_l$) stability of AQC operation independent of the behavior of the spectral density of the environment. 

Finally, we note that several approximations entered the above analysis, which is the reason this is a simplified analysis of thermal stability. Most importantly, we did not consider physical locality and weight restrictions on the EGP Hamiltonian and possible energy penalties. It would be interesting to extend this analysis to include physical restrictions and the structure of degenerate codes.

\end{document}